%% file: dcs.tex
\documentclass[11pt]{article}
\usepackage{widebar,ifpdf,graphicx,amsmath,amsbsy,amssymb,citesort,epsf,fullpage,url,subeqn}

\ifpdf

      \usepackage[pdftex]{epsfig}

      \usepackage[pdftex]{hyperref}

    \else

      \usepackage[dvips]{epsfig}

      \newcommand{\href}[2]{#2}

    \fi

\def\hatomega{ {\widehat{\Omega}} } 
\def\real    { \mathbb{R} }
\def \R {R} 
\def \j {j} 
\def \J {J} 
\def \m {m} 
\def \n {n} 
\def \N {N} 
\def \M {M} 
\def \K {K} 
\def \trans {^T} 
\def \numcols {D}
\def \Ebar {E}

\def \Etilde {\widebar{E}}
\newcommand \rank[1] {\mathrm{rank}( #1 )}

\newtheorem{THEO}{Theorem}
\newtheorem{LEMM}{Lemma}
\newtheorem{REMA}{Remark}
\newtheorem{DEFI}{Definition}

\newtheorem{CONJECTURE}{Conjecture}
\newtheorem{COROLLARY}{Corollary}
\newcommand{\cl}[1]{{\mathcal{#1}}}

\newcommand{\qed}{{\unskip\nobreak\hfil\penalty50\hskip2em\vadjust{}
           \nobreak\hfil$\Box$\parfillskip=0pt\finalhyphendemerits=0\par}}

\newcommand{\qq}{\vspace*{-2mm}}
\newcommand{\pp}{\vspace*{-2mm}}

\begin{document}
\renewcommand{\textfraction}{0}

\title{\LARGE\bf Distributed Compressive Sensing}

\author{\large \sl Dror Baron,$^{\textrm{1}}$  Marco F.\ Duarte,$^{\textrm{2}}$ Michael B.\ Wakin,$^{\textrm{3}}$
\\[3pt]
\large\sl Shriram Sarvotham,$^{\textrm{4}}$ and Richard G.\
Baraniuk$^{\textrm{2}}$$\:$\footnote{
This work was supported by the grants NSF CCF-0431150 and
CCF-0728867, 
DARPA HR0011-08-1-0078, DARPA/ONR N66001-08-1-2065, ONR N00014-07-1-0936 and
N00014-08-1-1112, AFOSR FA9550-07-1-0301, ARO MURI W311NF-07-1-0185,
and the Texas Instruments Leadership University Program.
Preliminary versions of this work have appeared at
the Allerton Conference on Communication, Control, and
Computing~\cite{DCSAllerton05}, the Asilomar Conference on
Signals, Systems and Computers~\cite{DCSAsilomar05}, the
Conference on Neural Information Processing Systems~\cite{DCSNIPS}, and the Workshop on Sensor, Signal and Information Processing~\cite{DCSSensip}.
\protect\\ E-mail: drorb@ee.technion.ac.il, \{duarte, shri, richb\}@rice.edu, mwakin@mines.edu;
Web: dsp.rice.edu/cs}\\[10pt]
\small $^{\textrm{1}}$Department of Electrical Engineering, Technion
-- Israel Institute of Technology, Haifa, Israel\\[-0pt]
\small $^{\textrm{2}}$Department of Electrical and Computer Engineering, Rice University, Houston, TX\\[-0pt]
\small $^{\textrm{3}}$Division of Engineering, Colorado School of Mines, Golden, CO\\[-0pt]
\small $^{\textrm{4}}$Halliburton, Houston, TX\\[-0pt]}
\date{}
\maketitle \thispagestyle{empty}
\vspace{-0.3in}

{\em This paper is dedicated to the memory of Hyeokho Choi, our colleague,
mentor, and friend.}

\begin{abstract}\noindent
Compressive sensing is a signal acquisition framework
based on the revelation that
a small collection of linear projections of a sparse signal contains
enough information for stable recovery. In this paper we introduce a
new theory for {\em distributed compressive sensing} (DCS) that
enables new distributed coding algorithms for multi-signal ensembles
that exploit both intra- and inter-signal correlation structures.
The DCS theory rests on a new concept that we term the {\em joint
sparsity} of a signal ensemble.
Our theoretical contribution is to characterize the fundamental
performance limits of DCS recovery for jointly sparse signal ensembles in the
noiseless measurement setting; our result connects single-signal,
joint, and distributed (multi-encoder) compressive sensing.
To demonstrate the efficacy of our framework and to show that
additional challenges such as computational tractability can be addressed,
we study in detail three example models for jointly sparse signals.
For these models, we develop practical algorithms for joint recovery of
multiple signals from incoherent projections. In two of our three models, 
the results are asymptotically best-possible, meaning that both
the upper and lower bounds match the performance of our practical
algorithms. Moreover, simulations indicate that the asymptotics take 
effect with just a moderate number of signals. DCS is immediately
applicable to a range of problems in sensor arrays and networks.
\end{abstract}
\normalsize


\section{Introduction}
\label{sec:intro}

A core tenet of signal processing and information theory is that
signals, images, and other data often contain some type of {\em
structure} that enables intelligent representation and processing.
The notion of structure has been characterized and exploited in a
variety of ways for a variety of purposes.  In this paper, we focus
on exploiting signal {\em correlations} for the purpose of {\em
compression}.

Current state-of-the-art compression algorithms employ a
decorrelating transform such as an exact or approximate
Karhunen-Lo\`{e}ve transform (KLT) to compact a correlated
signal's energy into just a few essential coefficients
\cite{devore92im,shapiro93em,xiong97sf}.  Such {\em transform
coders} exploit the fact that many signals have a {\em sparse}
representation in terms of some basis, meaning that a small number
$\K$ of adaptively chosen transform coefficients can be
transmitted or stored rather than $\N \gg \K$ signal samples. For
example, smooth signals are sparse in the Fourier basis, and
piecewise smooth signals are sparse in a wavelet basis
\cite{Mallat}; the commercial coding standards MP3~\cite{MP3},
JPEG~\cite{JPEG}, and JPEG2000~\cite{JPEG2000} directly exploit
this sparsity.

\qq
\subsection{Distributed source coding}
\qq

While the theory and practice of compression have been well
developed for individual signals, distributed sensing applications
involve multiple signals, for which there has been less progress.
Such settings are motivated by the proliferation of complex,
multi-signal acquisition architectures, such as acoustic and RF
sensor arrays, as well as sensor networks. These architectures 
sometimes involve battery-powered devices, which restrict the
communication energy, and high aggregate data rates, 
limiting bandwidth availability; both factors make the reduction 
of communication critical.

Fortunately, since the sensors presumably observe related phenomena,
the ensemble of signals they acquire can be expected to possess some
joint structure, or {\em inter-signal correlation}, in addition to
the {\em intra-signal correlation} within each individual sensor's
measurements. In such settings, {\em distributed source coding} that
exploits both intra- and
inter-signal correlations might allow the network to save on the
communication costs involved in exporting the ensemble of signals to
the collection point \cite{CoverThomas,SW73,DISCUS,Xiong}.
A number of distributed coding algorithms have been developed that
involve collaboration amongst the sensors~\cite{cristescu04,VetterliKL,Wag2005Jul5Distribute,Ortega2005}.
Note, however, that any collaboration involves some amount of inter-sensor
communication overhead.

In the {\em Slepian-Wolf} framework for lossless distributed
coding \cite{CoverThomas,SW73,DISCUS,Xiong}, the
availability of correlated side information at the decoder (collection point)
enables each sensor node to communicate losslessly
at its conditional entropy rate rather than at its individual
entropy rate, as long as the sum rate exceeds the joint entropy rate.
Slepian-Wolf coding has the distinct advantage that
the sensors need not collaborate while encoding their
measurements, which saves valuable communication overhead.
Unfortunately, however, most existing coding algorithms
\cite{DISCUS,Xiong} exploit only inter-signal correlations and not
intra-signal correlations. To date there has been only limited
progress on distributed coding of so-called ``sources with
memory.''
The direct implementation for sources with memory would require huge lookup
tables \cite{CoverThomas}. Furthermore, approaches combining pre- or
post-processing of the data to remove intra-signal correlations combined with
Slepian-Wolf coding for the inter-signal correlations appear to have limited
applicability, because such processing would alter the data in a way
that is unknown to other nodes.
Finally, although recent papers~\cite{uyematsu-universal,GF,ChenSW}
provide compression of spatially correlated sources with memory,
the solution is specific to lossless distributed compression and
cannot be readily extended to lossy compression settings.
We conclude that the design of constructive techniques for
distributed coding of sources with both intra- and inter-signal
correlation is a challenging problem with many potential applications.

\qq
\subsection{Compressive sensing (CS)}
\label{sec:cssingle}
\qq

A new framework for single-signal sensing and compression has developed
under the rubric of {\em compressive sensing}
(CS). CS builds on the work of Cand\`{e}s, Romberg,
and Tao \cite{CandesRUP} and Donoho~\cite{DonohoCS}, who showed that
if a signal has a sparse representation in one basis then it can be
recovered from a small number of projections onto a second basis
that is {\em incoherent} with the first.\footnote{Roughly speaking,
{\em incoherence} means that no element of one basis has a sparse
representation in terms of the other basis. This notion has a
variety of formalizations in the CS
literature~\cite{CandesCS,DonohoCS,TroppOMP}.}
CS relies on tractable recovery procedures that can
provide exact recovery of a signal of length $N$ and
sparsity $K$, i.e., a signal that can be written as a
sum of $\K$ basis functions from some known basis, where $\K$ can be
orders of magnitude less than $\N$. 

The implications of CS are promising for many applications,
especially for sensing signals that have a sparse representation in
some basis. Instead of sampling a $\K$-sparse signal $\N$ times,
only $M = O(\K \log N)$ incoherent measurements suffice, where $\K$ can be
orders of magnitude less than $\N$.
Moreover, the $M$ measurements need not be
manipulated in any way before being transmitted, except possibly
for some quantization.  Finally, independent and identically
distributed (i.i.d.)\ Gaussian or Bernoulli/Rademacher (random
$\pm 1$) vectors provide a useful {\em universal} basis that is
incoherent with all others.
Hence, when using a random basis, CS is universal in the sense that the
sensor can apply the same measurement mechanism no matter what
basis sparsifies the signal
\cite{jlcs}.

While powerful, the CS theory at present is designed mainly to
exploit intra-signal structures at a {\em single} sensor. In a multi-sensor setting,
one can naively obtain {\em separate measurements} from each signal and 
recover them separately. However, it is possible to obtain measurements that 
each depend on all signals in the ensemble by having sensors collaborate with each 
other in order to combine all of their measurements;
we term this process a {\em joint measurement setting}. In fact, initial work
in CS for multi-sensor settings used standard CS with joint measurement and recovery 
schemes that exploit inter-signal 
correlations~\cite{Nowak,Gossiping,Wang,saligrama2,saligrama3}. However, 
by recovering sequential time instances of the sensed data individually, these 
schemes ignore intra-signal correlations.

\qq
\subsection{Distributed compressive sensing (DCS)}
\qq

In this paper we introduce a new theory for {\em distributed
compressive sensing} (DCS) to enable new distributed coding
algorithms that exploit {\em both} intra- and inter-signal correlation
structures. In a typical DCS scenario, a number of sensors measure
signals that are each individually sparse in some basis and also 
correlated from sensor to sensor. Each sensor {\em separately} 
encodes its signal by projecting it onto another, incoherent basis 
(such as a random one) and then transmits just a few of the resulting 
coefficients to a single collection point. Unlike the joint measurement 
setting described in Section~\ref{sec:cssingle}, DCS requires no 
collaboration between the sensors during signal acquisition.
Nevertheless, we are able to exploit the inter-signal correlation by
using all of the obtained measurements to recover all the signals 
{\em simultaneously}.  Under the right conditions, a decoder at the 
collection point can recover each of the signals precisely.

The DCS theory rests on a concept that we term the {\em joint
sparsity} --- the sparsity of the entire signal ensemble.
The joint sparsity is often smaller than the aggregate over individual
signal sparsities. Therefore, DCS offers a reduction
in the number of measurements, in a manner analogous to the rate reduction
offered by the Slepian-Wolf framework~\cite{SW73}. Unlike the single-signal
definition of sparsity, however, there are numerous plausible ways in which
joint sparsity could be defined. In this paper, we first provide a general
framework for joint sparsity using algebraic formulations based on a
graphical model. Using this framework, we derive bounds
for the number of measurements necessary for recovery under
a given signal ensemble model. Similar to Slepian-Wolf coding~\cite{SW73}, 
the number of measurements required for each sensor must account for the 
minimal features unique to that sensor, while at the same time features that 
appear among multiple sensors must be amortized over the group. Our bounds
are dependent on the dimensionality of the subspaces in which each group of
signals reside; they afford a reduction in the number of measurements that we
quantify through the notions of {\em joint} and {\em conditional}
sparsity, which are conceptually related to joint and conditional entropies.
The common thread is that dimensionality and entropy both quantify the volume
that the measurement and coding rates must cover.
Our results are also applicable to cases where the signal ensembles
are measured jointly, as well as to the single-signal case.

While our general framework does not by design provide insights for 
computationally efficient recovery, we also provide interesting models 
for joint sparsity where our results carry through from the general 
framework to realistic settings with low-complexity algorithms.
In the first model, each signal is itself sparse, and so we could use CS to 
separately encode and decode each signal. However, there also exists a 
framework wherein a joint sparsity model for the ensemble uses fewer 
total coefficients. In the second model, all signals share the locations of the 
nonzero coefficients. In the third model, no signal is itself sparse, yet there still 
exists a joint sparsity among the signals that allows recovery from
significantly fewer than $N$ measurements per sensor.
For each model we propose tractable algorithms for joint signal recovery, 
followed by theoretical and
empirical characterizations of the number of measurements per sensor 
required for accurate recovery. We show that, under these models, joint 
signal recovery can recover signal ensembles from significantly fewer 
measurements than would be required to recover each signal individually. 
In fact, for two of our three models we obtain best-possible performance 
that could not be bettered by an oracle that knew the the indices of the nonzero
entries of the signals.

This paper focuses primarily on the basic task of reducing the number 
of measurements for recovery of a signal ensemble in order to reduce the
communication cost of source coding that ensemble. Our emphasis is on 
noiseless measurements of strictly sparse signals, where the optimal 
recovery relies on $\ell_0$-norm optimization,\footnote{The $\ell_0$ 
``norm'' $\|x\|_0$ merely counts the number of nonzero entries in the vector 
$x$.} which is computationally intractable. In practical settings, additional 
criteria may be relevant for measuring performance. For example, the 
measurements will typically be real numbers that must be quantized,
which gradually degrades the recovery quality as the quantization 
becomes coarser \cite{CandesDS,bitscs}. Characterizing DCS in
light of practical considerations such as rate-distortion tradeoffs, power
consumption in sensor networks, etc., are topics of future
research~\cite{saligrama2,saligrama3}.

\qq
\subsection{Paper organization}
\qq

Section~\ref{sec-cs} overviews the single-signal CS theories and provides
a new result on CS recovery. While some readers may be familiar with this
material, we include it to make the paper self-contained.
Section~\ref{sec-models} introduces our general framework for joint sparsity
models and proposes three example models for joint
sparsity. We provide our detailed analysis for the general framework in
Section~\ref{sec-theory}; we then address the three models in
Section~\ref{sec-algs}.  We close the paper with a discussion and conclusions in
Section~\ref{sec-discussion}.  Several appendices contain the proofs.

\qq
\section{Compressive Sensing Background} \label{sec-cs}
\qq

\subsection{Transform coding} \qq

Consider a real-valued signal\footnote{Without loss of generality, we will focus
on one-dimensional signals (vectors) for
notational simplicity; the extension to multi-dimensional signal, e.g., images, is
straightforward.} $x\in \real^N$ indexed as $x(\n)$, $\n \in
\{1,2,\dots,\N\}$. Suppose that the basis $\Psi =
[\psi_1,\ldots,\psi_\N]$ provides a $\K$-sparse
representation of $x$; that is,
\begin{equation}
x = \sum_{\n=1}^\N \vartheta(\n)\, \psi_\n = \sum_{k=1}^K
\vartheta({\n_k}) \, \psi_{\n_k},
\label{eq:tcsignal}
\end{equation}
where $x$ is a linear combination of $\K$ vectors chosen from
$\Psi$, $\{\n_k\}$ are the indices of those vectors, and
$\{\vartheta(\n)\}$ are the coefficients; the concept is extendable
to tight frames \cite{Mallat}. Alternatively, we can write in
matrix notation
$x = \Psi \vartheta$,
where $x$ is an $\N\times 1$ column vector, the {\em sparse basis}
matrix $\Psi$ is $\N\times\N$ with the basis vectors $\psi_\n$ as
columns, and $\vartheta$ is an $\N\times 1$ column vector with $\K$
nonzero elements. Using $\|\cdot\|_p$ to denote the $\ell_p$
norm, we can write
that $\|\vartheta\|_0=\K$; we can also write the set of nonzero indices
$\Omega \subseteq \{1,\ldots,N\}$, with $|\Omega| = K$. Various
expansions, including
wavelets~\cite{Mallat}, Gabor bases~\cite{Mallat},
curvelets~\cite{Curvelets}, etc., are widely used for representation
and compression of natural signals, images, and other data.

The standard procedure for compressing sparse and nearly-sparse signals, 
known as {\em transform coding}, is to ({\em i})~acquire the full $\N$-sample
signal $x$; ({\em ii})~compute the complete set of transform
coefficients $\{\vartheta(\n)\}$; ({\em iii})~locate the $\K$ largest,
significant coefficients and discard the (many) small coefficients;
({\em iv})~encode the {\em values and locations} of the largest
coefficients.
This procedure has three inherent inefficiencies: First, for a
high-dimensional signal, we must start with a large number of
samples $\N$. Second, the encoder must compute {\em all} of the $\N$
transform coefficients $\{\vartheta(\n)\}$, even though it will discard
all but $\K$ of them. Third, the encoder must encode the locations
of the large coefficients, which requires increasing the coding rate
since the locations change with each signal.

We will focus our theoretical development on exactly 
$\K$-sparse signals and
defer discussion of the more general situation of {\em compressible} signals 
where the coefficients decay rapidly with a power law but not to zero. Section
\ref{sec-discussion} contains additional discussion on real-world compressible
signals, and \cite{DCSIPSN06} presents simulation results.

\qq
\subsection{Incoherent projections} \qq

These inefficiencies raise a simple question: For a given signal, is it possible
to directly estimate the set of large $\vartheta(\n)$'s that will not
be discarded?  While this seems improbable, Cand\`{e}s, Romberg,
and Tao~\cite{CandesRUP,CandesCS} and Donoho \cite{DonohoCS} have
shown that a reduced set of projections can contain enough
information to recover sparse signals. A framework to acquire sparse signals,
often referred to as {\em compressive sensing} (CS)
\cite{CSweb}, has emerged that builds on this principle.

In CS, we do not measure or encode the $\K$ significant $\vartheta(\n)$
directly. Rather, we measure and encode $\M< N$ projections $y(\m) =
\langle x, \phi_\m\trans \rangle$ of the signal onto a {\em second
set} of functions $\{\phi_\m\},\m=1,2,\ldots,\M$, where
$\phi_\m\trans$ denotes the transpose of $\phi_\m$ and $\langle
\cdot,\cdot \rangle$ denotes the inner product. In matrix notation,
we measure
$y = \Phi x$,
where $y$ is an $\M\times 1$ column vector and the {\em measurement
matrix} $\Phi$ is $\M\times \N$ with each row a {\em measurement vector}
$\phi_\m$. Since $\M < \N$, recovery of the signal $x$ from the
measurements $y$ is ill-posed in general; however the additional
assumption of signal {\em sparsity} makes recovery possible and
practical.

The CS theory tells us that when certain conditions hold, namely
that the basis $\{\psi_\n\}$ cannot sparsely represent the
vectors $\{\phi_\m\}$ (a condition known as {\em
incoherence} \cite{CandesCS,DonohoCS,TroppOMP}) and the number of
measurements $\M$ is large enough (proportional to $K$), then it is indeed
possible to recover the set of large $\{\vartheta(\n)\}$ (and thus the signal
$x$) from the set of measurements $\{y(\m)\}$ \cite{CandesCS,DonohoCS}. This
incoherence property holds for many pairs of bases, including for
example, delta spikes and the sine waves of a Fourier basis, or
the Fourier basis and wavelets. Signals that are sparsely
represented in frames or unions of bases can be recovered from
incoherent measurements in the same fashion. Significantly, this incoherence
also holds with high probability between {\em any} arbitrary fixed basis or frame
and a randomly generated one. In the sequel, we will focus our analysis
to such {\em random measurement} procedures.

\qq
\subsection{Signal recovery via $\ell_0$-norm minimization}
\label{subsec-CSrecovery} \qq

The recovery of the sparse set of significant coefficients
$\{\vartheta(\n)\}$ can be achieved using {\em optimization} by
searching for the signal with the sparsest coefficient vector
$\{\widehat{\vartheta}(\n)\}$ that agrees with the $\M$ observed measurements in
$y$ (recall that $\M < \N$). Recovery relies on the key
observation that, under mild conditions on $\Phi$ and
$\Psi$, the coefficient vector $\vartheta$ is the unique solution to the
$\ell_0$-norm minimization
\begin{equation}
\widehat{\vartheta} = \arg\min \|\vartheta\|_0 ~~~\mbox{s.t. }
y=\Phi\Psi\vartheta
\label{eq:L0}
\end{equation}
with overwhelming probability. (Thanks to the incoherence between
the two bases, if the original signal is sparse in the $\vartheta$
coefficients, then no other set of sparse signal coefficients
$\vartheta^\prime$ can yield the same projections $y$.)

In principle, remarkably few incoherent measurements are required
to recover a $\K$-sparse signal via $\ell_0$-norm minimization.
More than $\K$ measurements must be taken to avoid
ambiguity; the following theorem, proven in Appendix~\ref{ap:kplusone},
establishes that $\K+1$ random
measurements will suffice. Similar results were established by
Venkataramani and Bresler~\cite{VenBres98}.

\begin{THEO}
\label{theo:kplusone} Let $\Psi$ be an orthonormal basis for
$\mathbb{R}^\N$, and let $1 \le \K < \N$. Then:
\begin{enumerate}
\item Let $\Phi$ be an $\M \times \N$ measurement matrix with
i.i.d.\ Gaussian entries with $\M \ge 2\K$. Then all signals $x = \Psi\vartheta$
having expansion coefficients $\vartheta \in \mathbb{R}^\N$ that
satisfy $\|\vartheta\|_0 = \K$ can be recovered uniquely from the
$\M$-dimensional measurement vector $y = \Phi x$ via the $\ell_0$-norm
minimization (\ref{eq:L0}) with probability one over $\Phi$.
\item Let $x = \Psi\vartheta$ such that
$\|\vartheta\|_0 = \K$. Let $\Phi$ be an $\M \times \N$ measurement
matrix with i.i.d.\ Gaussian entries (notably, independent of $x$)
with $\M \ge \K+1$.  Then $x$ can be recovered uniquely from the
$\M$-dimensional measurement vector $y = \Phi x$ via the $\ell_0$-norm
minimization (\ref{eq:L0}) with probability one over $\Phi$.
\item Let $\Phi$ be an $\M \times \N$
measurement matrix, where $\M \le \K$. Then, aside from
pathological cases (specified in the proof), no signal $x =
\Psi\vartheta$ with $\|\vartheta\|_0 = \K$ can be uniquely recovered
from the $\M$-dimensional measurement vector $y = \Phi x$.
\end{enumerate}
\end{THEO}

\begin{REMA}
The second statement of the theorem differs from the first in the
following respect: when $\K < \M < 2\K$, there will necessarily
exist $\K$-sparse signals $x$ that cannot be uniquely recovered
from the $\M$-dimensional measurement vector $y = \Phi x$.
However, these signals form a set of measure zero within the set
of {\em all} $\K$-sparse signals and can safely be avoided with 
high probability if $\Phi$ is randomly generated independently of $x$.
\end{REMA}

Comparing the second and third statements of
Theorem~\ref{theo:kplusone}, we see that one measurement separates the
{\em achievable region}, where perfect recovery is possible
with probability one, from the {\em converse region}, where with
overwhelming probability recovery is impossible. Moreover,
Theorem~\ref{theo:kplusone} provides a {\em strong converse
measurement region} in a manner analogous to the strong channel
coding converse theorems of information theory~\cite{CoverThomas}.

Unfortunately, solving the $\ell_0$-norm minimization problem is
prohibitively complex, requiring a combinatorial enumeration of the
${\N}\choose{\K}$ possible sparse subspaces.
In fact, the $\ell_0$-norm minimization problem in general is known to be
NP-hard~\cite{CandesECLP}. Yet another challenge is robustness;
in the setting of Theorem~\ref{theo:kplusone}, the recovery may be
very poorly conditioned. In fact, {\em both} of these considerations
(computational complexity and robustness) can be addressed, but at
the expense of slightly more measurements.

\qq
\subsection{Signal recovery via $\ell_1$-norm minimization}
\label{subsec-CSrecovery2} \qq

The practical revelation that supports the new CS theory is that it
is not necessary to solve the $\ell_0$-norm minimization to
recover the set of significant $\{\vartheta(\n)\}$. In fact, a much
easier problem yields an equivalent solution (thanks again to the
incoherence of the bases); we need only solve for the smallest
$\ell_1$-norm coefficient vector $\vartheta$ that agrees with the
measurements $y$
\cite{CandesCS,DonohoCS}:
\begin{equation}
\widehat{\vartheta} = \arg\min \|\vartheta\|_1 ~~~\mbox{s.t. }
y=\Phi\Psi\vartheta.
\label{eq:L1}
\end{equation}
This optimization problem, also known as {\em Basis
Pursuit}, is significantly more approachable and
can be solved with traditional linear programming techniques whose
computational complexities are polynomial in $\N$.

There is no free lunch, however; according to the theory, more
than $\K+1$ measurements are required in order to recover sparse
signals via Basis Pursuit. Instead, one typically requires $\M \ge
c\K$ measurements, where $c > 1$ is an {\em overmeasuring factor}.
As an example, we quote a result asymptotic in $\N$. For
simplicity, we assume that the sparsity scales linearly with $\N$;
that is, $\K = S\N$, where we call $S$ the {\em sparsity rate}.

\begin{THEO} {\rm \cite{CandesECLP,DonohoMar2005,DonohoJan2005}}
\label{theo:bpconstant} Set $\K = S\N$ with $0 < S \ll 1$. Then
there exists an overmeasuring factor $c(S) = O(\log(1/S))$, $c(S)
> 1$, such that, for a $\K$-sparse signal $x$ in basis $\Psi$, the
following statements hold:
\begin{enumerate}

\item The probability of recovering $x$ via $\ell_1$-norm minimization from
$(c(S) + \epsilon)\K$ random projections, $\epsilon > 0$,
converges to one as $\N \rightarrow \infty$.

\item The probability of recovering $x$ via $\ell_1$-norm minimization from
$(c(S) - \epsilon)\K$ random projections, $\epsilon > 0$,
converges to zero as $\N \rightarrow \infty$.
\end{enumerate}
\end{THEO}

In an illuminating series of papers, Donoho and Tanner
\cite{DonohoMar2005,DonohoJan2005,DonohoCountingFaces} have
characterized the overmeasuring factor $c(S)$ precisely. In our
work, we have noticed that the overmeasuring factor is quite similar
to $\log_2(1+S^{-1})$. We find this expression a useful rule of
thumb to approximate the precise overmeasuring ratio. Additional overmeasuring is
proven to provide robustness to measurement noise and quantization error \cite{CandesCS}.

Throughout this paper we use the abbreviated
notation $c$ to describe the overmeasuring factor required in
various settings even though $c(S)$ depends on the sparsity $\K$
and signal length $\N$.

\qq
\subsection{Signal recovery via greedy pursuit} \qq

Iterative greedy algorithms have also been developed to recover the signal
$x$ from the measurements $y$.
The Orthogonal Matching Pursuit (OMP) algorithm, for example, iteratively
selects the vectors from the matrix $\Phi\Psi$ that contain most of the energy of
the measurement vector $y$.  The selection at each iteration is made
based on inner products between the columns of $\Phi\Psi$ and a
residual; the residual reflects the component of $y$ that is
orthogonal to the previously selected columns. The algorithm has been proven to
successfully recover the acquired signal from incoherent measurements
with high  probability, at the expense of slightly more
measurements,~\cite{TroppOMP,CDDNOA}.
Algorithms inspired by OMP, such as regularized orthogonal matching pursuit~\cite{ROMP}, CoSaMP~\cite{cosamp}, and Subspace Pursuit~\cite{SP} have
been shown  to attain similar guarantees to those of their optimization-based
counterparts. In the following, we will exploit both Basis Pursuit and greedy
algorithms for
recovering jointly sparse signals from incoherent measurements.

\qq
\subsection{Properties of random measurements}

In addition to offering substantially reduced measurement rates, CS has
many attractive and intriguing properties, particularly when
we employ random projections at the sensors.  Random measurements
are {\em universal} in the sense that any sparse basis can be used,
allowing the same encoding strategy to be applied in different sensing
environments. Random measurements are also {\em future-proof}: if a
better sparsity-inducing basis is found for the signals, then the same
measurements can be used to recover a more accurate view of the
environment. Random coding is also {\em robust}: the measurements
coming from each sensor have equal priority, unlike Fourier or wavelet
coefficients in current coders. Finally, random measurements allow a 
{\em progressively better recovery} of the data as more measurements 
are obtained; one or more measurements can also be lost without 
corrupting the entire recovery.

\qq
\subsection{Related work} \label{subsub-nowak} \qq

Several researchers have formulated {\em joint measurement settings} for CS in sensor networks that exploit
inter-signal correlations~\cite{Nowak,Gossiping,Wang,saligrama2,saligrama3}.
In their approaches, each sensor $\n \in \{1,2,\ldots,\N\}$ simultaneously
records a single reading $x(\n)$ of some spatial field (temperature
at a certain time, for example).\footnote{Note that in Section~\ref{subsub-nowak}
only, $\N$ refers to the number of sensors, since each sensor acquires a signal sample.}
Each of the sensors generates a pseudorandom sequence
$r_\n(\m), \m=1,2,\ldots,\M$, and modulates the reading as $x(\n)
r_\n(\m)$. Each sensor $\n$ then transmits its $\M$ numbers in
sequence to the collection point
where the measurements are aggregated, obtaining $\M$
measurements $y(m) = \sum_{n=1}^N x(\n)r_\n(\m)$. Thus, defining $x
= [x(1),x(2),\ldots,x(\N)]\trans$ and $\phi_\m =
[r_1(\m),r_2(\m),\ldots,r_\N(\m)]$, the collection point
automatically receives the measurement vector
$y=[y(1),y(2),\ldots,y(\M)]\trans = \Phi x$ after $\M$ transmission
steps. The samples $x(\n)$ of the spatial field can then be
recovered using CS provided that $x$ has a sparse representation in
a known basis.
These methods have a major limitation: since they operate at a single time
instant, they exploit only inter-signal and not intra-signal correlations;
that is, they essentially assume that the sensor field is i.i.d.\
from time instant to time instant. In contrast, we will develop
signal models and algorithms that are agnostic to the spatial
sampling structure and that exploit both inter- {\em and} intra-signal correlations.

Recent work has adapted DCS to the finite rate of innovation signal
acquisition framework~\cite{DCSFROI} and to the continuous-time
setting~\cite{EldarMMV}. Since the original submission of this paper, 
additional work has focused on the analysis and
proposal of recovery algorithms for jointly sparse signals~\cite{jsconst,atoms}.

\qq
\section{Joint Sparsity Signal Models} \label{sec-models}
\qq

In this section, we generalize the notion of a signal being sparse
in some basis to the notion of an ensemble of signals being {\em
jointly sparse}.

\subsection{Notation}

We will use the following notation for signal ensembles and our
measurement model. Let $\Lambda:=\{1,2,\dots,\J\}$ denote the
set of indices for the $J$ signals in the ensemble. Denote the {\em signals} in the
ensemble by
$x_\j$, with $\j \in \Lambda$ and assume that each
signal $x_\j \in \mathbb{R}^\N$. We use $x_\j(\n)$ to denote sample
$\n$ in signal $\j$, and assume for the sake of illustration --- but without loss of
generality ---  that these signals are sparse in the canonical basis, 
i.e., $\Psi= \mathbf{I}$.
The entries of the signal can take arbitrary real values.

We denote by $\Phi_\j$ the measurement matrix for signal $\j$;
$\Phi_\j$ is $M_\j \times \N$ and, in general, the entries of
$\Phi_\j$ are different for each $\j$. Thus, $y_\j = \Phi_\j x_\j$
consists of $\M_\j < \N$ random measurements of $x_\j$.
We will emphasize random i.i.d.\ Gaussian matrices $\Phi_\j$ in the
following, but other schemes are possible, including random $\pm 1$
Bernoulli/Rademacher matrices, and so on.

To compactly represent the signal and measurement ensembles, we denote
$\widebar{M} = \sum_{j\in\Lambda} M_j$
and
define $X \in \mathbb{R}^{JN}$, $Y \in \mathbb{R}^{\widebar{M}}$, and
$\Phi \in \mathbb{R}^{\widebar{M} \times JN}$ as
\begin{equation}
X = \left[
\begin{array}{ccc}
x_1 \\
x_2 \\
\vdots \\
x_J \end{array} \right], ~~~~~ Y = \left[
\begin{array}{ccc}
y_1 \\
y_2 \\
\vdots \\
y_J \end{array} \right],  \mbox{~~~~~ and ~~~~~} \Phi = \left[
\begin{array}{cccc}
\Phi_1 & {\bf 0} & \hdots & {\bf 0}\\
 {\bf 0} & \Phi_2 & \hdots & {\bf 0}\\
\vdots &\vdots&\ddots&\vdots\\
 {\bf 0}&{\bf 0}&\hdots& \Phi_J \end{array} \right],
\label{eq:phimatrix}
\end{equation}
with {\bf 0} denoting a matrix of appropriate size with all entries
equal to 0. We then have $Y = \Phi X$. Equation~(\ref{eq:phimatrix}) 
shows that separate measurement matrices have a characteristic 
block-diagonal structure when the entries of the sparse vector are 
grouped by signal.

Below we propose a general framework for {\em joint sparsity models}
(JSMs) and three example JSMs that apply in different
situations.

\subsection{General framework for joint sparsity}
\label{sec:generalframework}

We now propose a general framework to quantify the sparsity of an ensemble of
correlated signals $x_1,
x_2,\ldots, x_J$, which allows us to compare
the complexities of different signal ensembles and to quantify their measurement requirements.
The framework is based on a factored
representation of the signal ensemble that decouples its location
and value information.

To motivate this factored representation, we begin by examining the structure
of a single sparse signal, where $x \in \real^N$ with $K \ll N$
nonzero entries. As an alternative to the notation used in (\ref{eq:tcsignal}),
we can decouple the location and value information in $x$ by writing
$x = P \theta$, where $\theta \in \real^K$ contains only
the nonzero entries of $x$, and $P$ is an {\em identity submatrix},
i.e., $P$ contains $K$ columns of
the $N \times N$ identity matrix $\mathbf{I}$.
Any $K$-sparse signal can be written in similar fashion.
To model the set of all possible sparse
signals, we can then let $\mathcal{P}$ be the set of all
identity submatrices of all possible sizes $N \times K'$, with $1
\le K' \le N$. We refer to $\mathcal{P}$ as a {\em sparsity model}.
Whether a signal is sufficiently sparse is defined {\em in the context of
this model}: given a signal $x$, one can consider all possible
factorizations $x = P \theta$ with $P \in \mathcal{P}$.
Among these factorizations, the unique representation with smallest 
dimensionality for $\theta$ equals the {\em sparsity level} of the 
signal $x$ under the model $\mathcal{P}$.

In the signal ensemble case, we consider factorizations of the form
$X = P\Theta$ where $X \in \real^{JN}$ as above, $P \in \real^{JN
\times \delta}$, and $\Theta \in \real^\delta$ for various integers $\delta$. 
We refer to $P$ and $\Theta$
as the {\em location matrix} and {\em value vector}, respectively. A
{\em joint sparsity model} (JSM) is defined in terms of a set
$\mathcal{P}$ of admissible location matrices $P$ with varying
numbers of columns; we specify below additional conditions that the matrices $P$
must satisfy for each model.
For a given ensemble $X$, we let $\mathcal{P}_F(X) \subseteq \mathcal{P}$ denote
the set of feasible location matrices $P \in \mathcal{P}$ for which a
factorization $X = P\Theta$ exists. We define the {\em joint sparsity level}
of the signal ensemble as follows.

\begin{DEFI}
\label{def:joint_spars}
The {\em joint sparsity level} $D$ of the signal ensemble $X$ is the number of columns
of the smallest 
matrix $P \in \mathcal{P}_F(X)$.
\end{DEFI}

In contrast to the single-signal case, there are several
natural choices for what matrices $P$ should be members of a joint
sparsity model $\mathcal{P}$. We restrict our attention in the sequel to what we
call {\em common/innovation component JSMs}.  In these models each signal
$x_j$ is generated as a combination of two components: ($i$) a common
component $z_C$, which is present in all signals, and ($ii$) an
innovation component $z_j$, which is unique to each signal. These
combine additively, giving
$$ x_j = z_C+z_j,~~j \in \Lambda. $$
Note, however, that the individual components might be zero-valued in specific scenarios. We can express the component signals as
\begin{equation*}
z_C = P_C\theta_C,~~z_\j = P_\j\theta_j, ~~~ \j \in \Lambda,
\end{equation*}
where $\theta_C \in \real^{K_C}$ and each $\theta_j \in \real^{K_j}$ have nonzero entries.
Each matrix $P \in \mathcal{P}$ that can express such signals $\{x_j\}$ has the form
\begin{equation}
P = \left[
\begin{array}{ccccc}
P_C & P_1 & {\bf 0} & \hdots & {\bf 0}\\
P_C & {\bf 0} & P_2 & \hdots & {\bf 0}\\
\vdots & \vdots &\vdots&\ddots&\vdots\\
P_C & {\bf 0}& {\bf 0} &\hdots& P_J \end{array} \right],
\label{eq:pmatrix}
\end{equation}
where $P_C$, $\{P_\j\}_{j \in \Lambda}$ are identity
submatrices.
We define the value vector as
$\Theta = [\theta_C^T~\theta_1^T~\theta_2^T~\ldots~\theta_J^T]^T$,
where $\theta_C \in \real^{K_C}$ and each $\theta_j \in
\real^{K_j}$, to obtain $X = P \Theta$. Although the values of $K_C$ and $K_j$ are dependent on the matrix $P$, we omit this dependency in the sequel for brevity, except when necessary for clarity.

If a signal ensemble $X = P \Theta$,
$\Theta \in \real^{\delta}$ were to be generated by a selection of $P_C$
and $\{P_j\}_{j \in \Lambda}$, where all $J+1$ identity
submatrices share a common column vector, then
$P$ would not be full rank. 
In other cases, we may observe a vector $\Theta$ that has zero-valued entries; i.e., we may have $\theta_j(k) = 0$ for some $1 \le k \le K_j$ and some $j \in \Lambda$, or $\theta_C(k) = 0$ for some $1 \le k \le K_C$.
In both of these cases, by removing
one instance of this column from any of the identity submatrices,
one can obtain a matrix $Q$ with fewer columns 
for which there exists $\Theta' \in \real^{\delta-1}$ that gives $X = Q \Theta'$. If $Q \in \mathcal{P}$,
then we term this phenomenon {\em sparsity reduction}. Sparsity reduction, when
present, reduces the effective joint sparsity of a signal ensemble.
As an example of sparsity reduction, consider $\J=2$ signals of length $N=2$. 
Consider the coefficient $z_C(1) \ne 0$ of the common component $z_C$ and the
corresponding innovation coefficients $z_1(1), z_2(1) \ne 0$.
Suppose that all other coefficients are zero.
The location matrix $P$ that arises is
$$
P = \left[
\begin{array}{ccccc}
1 & 1 & 0\\
0 & 0 & 0\\
1 & 0 & 1\\
0 & 0 & 0
\end{array} \right].
$$
The span of this location matrix (i.e., the set of signal ensembles $X$
that it can generate) remains unchanged if we remove any one of the columns,
i.e., if we drop any entry of the value vector $\Theta$. This provides us with
a lower-dimensional representation $\Theta'$ of the same signal ensemble
$X$ under the JSM $\mathcal{P}$; the joint sparsity of $X$ is $D=2$.

\qq
\subsection{Example joint sparsity models}
\label{sec:exjsms}
\pp

Since different real-world scenarios lead to different forms of correlation 
within an ensemble of sparse signals, we consider
several possible designs for a JSM $\mathcal{P}$.
The distinctions among our three JSMs concern the differing sparsity
assumptions regarding the common and innovation components.

\subsubsection{JSM-1: Sparse common component + innovations}
\label{sec-ds1} \qq

In this model, we suppose that each signal contains a common component 
$z_C$ that is {\em sparse} plus an innovation component $z_j$ that is also {\em
sparse}. Thus, this joint sparsity model (JSM-1)
$\mathcal{P}$ is represented by the set of all matrices of the form
(\ref{eq:pmatrix}) with $K_C$ and all $K_j$ smaller than $N$. Assuming 
that sparsity reduction is not possible, the joint sparsity $D = K_C +
\sum_{j \in \Lambda}K_j$.

A practical situation well-modeled by this framework is a
group of sensors measuring temperatures at a number of outdoor
locations throughout the day. The temperature readings $x_\j$ have
both temporal (intra-signal) and spatial (inter-signal) correlations.
Global factors, such as the sun and
prevailing winds, could have an effect $z_C$ that is both common to
all sensors and structured enough to permit sparse representation.
More local factors, such as shade, water, or animals, could
contribute localized innovations $z_\j$ that are also structured
(and hence sparse). A similar scenario could be imagined for a
network of sensors recording light intensities, air pressure, or
other phenomena. All of these scenarios correspond to measuring
properties of physical processes that change smoothly in time and in
space and thus are highly correlated~\cite{estrin_2002,pottie:cacm00}.

\qq
\subsubsection{JSM-2: Common sparse supports}
\label{sec-ds2} \qq

In this model, the common component $z_C$ is equal
to zero, each innovation component $z_j$ is {\em sparse}, and
the innovations $\{z_j\}$ share the {\em same sparse support}
but have different nonzero coefficients.
To formalize this setting in a joint sparsity model (JSM-2) we let
$\mathcal{P}$ represent the set of all matrices of the form (\ref{eq:pmatrix}),
where $P_C = \emptyset$ and $P_j = \widebar{P}$ for all $j \in \Lambda$. Here 
$\widebar{P}$
denotes an arbitrary identity submatrix of size $N\times K$, with $K \ll N$. For a given $X = P \Theta$, we may again partition the value vector
$\Theta = [\theta_1^T~\theta_2^T~\ldots~\theta_J^T]^T$,
where each $\theta_j \in \real^{K}$. 
It is easy to see that the matrices $P$ from JSM-2 are full rank. 
Therefore, when sparsity reduction is not possible, the joint
sparsity $D = JK$.

The JSM-2 model is immediately applicable to acoustic and RF sensor arrays,
where each sensor acquires
a replica of the same Fourier-sparse signal but with phase shifts and
attenuations caused by signal propagation. In this case, it is
critical to recover each one of the sensed signals. Another
useful application for this framework is MIMO communication
\cite{Tropp05}.

Similar signal models have been considered in the area of
{\em simultaneous sparse
approximation}~\cite{Tropp05,TemlyakovSA,Cotter05}. In this setting,
a collection of sparse signals share the same expansion vectors from
a redundant dictionary. The sparse approximation can be recovered
via greedy algorithms such as {\em Simultaneous Orthogonal Matching
Pursuit} (SOMP)~\cite{Tropp05,TemlyakovSA} or {\em MMV Order
Recursive Matching Pursuit} (M-ORMP)~\cite{Cotter05}. We use the
SOMP algorithm in our setting (Section~\ref{sec-mrr2}) to
recover from incoherent measurements an ensemble of signals sharing
a common sparse structure.

\qq
\subsubsection{JSM-3: Nonsparse common component + sparse innovations}
\label{sec-ds3} \qq

In this model, we suppose that each signal contains an arbitrary common 
component $z_C$ and a sparse innovation component $z_j$; this model 
extends JSM-1 by relaxing the assumption that the common component 
$z_C$ has a sparse representation.
To formalize this setting in the JSM-3 model, we let
$\mathcal{P}$ represent the set of all matrices (\ref{eq:pmatrix}) in which
$P_C = I$, the $N \times N$ identity matrix. This implies each $K_j$ is
smaller than $N$ while $K_C = N$; thus, we obtain $\theta_C \in \real^{N}$
and $\theta_j \in \real^{K_j}$.
Assuming that sparsity reduction is not possible, the joint sparsity $D = N +
\sum_{j \in \Lambda}K_j$.
We also consider the specific case where the supports of the
innovations are shared by all signals, which extends JSM-2; in this
case we will have $P_j = \widebar{P}$ for all $j \in \Lambda$, with $\widebar{P}$ an
identity submatrix of size $N\times K$. It is easy to see that in this case
sparsity reduction is possible, and so the the joint sparsity can drop to
$D = N+(J-1)K$.
Note that {\em separate} CS recovery is impossible in
JSM-3 with any fewer than $N$ measurements per sensor, since the
common component is not sparse. However, we will demonstrate that
{\em joint} CS recovery can indeed exploit the common
structure.

A practical situation well-modeled by this framework is where several
sources are recorded by different sensors together with a background
signal that is not sparse in any basis. Consider, for example, a verification system in a
component production plant, where cameras acquire snapshots of each
component to check for manufacturing defects.  While each image could be
extremely complicated, and hence nonsparse, the ensemble of images will be highly
correlated, since each camera is observing the same device with
minor (sparse) variations.

JSM-3 can also be applied in non-distributed scenarios. For
example, it motivates the compression of data such as video, where
the innovations or differences between video frames may be sparse,
even though a single frame may not be very sparse. In this case,
JSM-3 suggests that we encode each video frame separately using
CS and then decode all frames of the video sequence jointly.  This
has the advantage of moving the bulk of the computational complexity
to the video decoder. The PRISM system proposes a similar
scheme based on Wyner-Ziv distributed encoding~\cite{PRISMAllerton02}.

There are many possible joint sparsity models beyond those 
introduced above, as well as beyond the common and innovation component 
signal model. Further work will yield new JSMs suitable for other application 
scenarios; an example application consists of multiple cameras taking digital 
photos of a common scene from various angles \cite{Wagner03}. Extensions 
are discussed in Section~\ref{sec-discussion}.

\qq
\section{Theoretical Bounds on Measurement Rates} \qq
\label{sec-theory}

In this section, we seek conditions on $\mathcal{M} = ( M_1, M_2, \ldots, M_J )$, the tuple
of number of measurements from each sensor, such that we can
guarantee perfect recovery of $X$ given $Y$.
To this end, we provide a graphical model for the general framework provided in
Section~\ref{sec:generalframework}. This graphical model is fundamental in the
derivation of the number of measurements needed for each sensor, as well
as in the formulation of a combinatorial recovery procedure.
Thus, we generalize Theorem~\ref{theo:kplusone} to
the distributed setting to obtain fundamental limits on the number of
measurements that enable recovery of sparse signal ensembles.

Based on the models presented in Section~\ref{sec-models},
recovering $X$ requires determining a value vector $\Theta$ and location
matrix $P$ such that $X=P\Theta$.
Two challenges immediately present themselves.
First, a given measurement depends only on some of the
components of $\Theta$, and the measurement budget should be adjusted
between the sensors according to the information that can be gathered on 
the components of $\Theta$.
For example, if a component $\Theta(d)$ does not affect any signal
coefficient $x_j(\cdot)$ in sensor $j$, then the corresponding measurements
$y_j$ provide no information about $\Theta(d)$.
Second, the decoder must identify a location matrix
$P \in \mathcal{P}_F(X)$ from the set $\mathcal{P}$ and the measurements $Y$.

\qq
\subsection{Modeling dependencies using bipartite graphs}
\label{sec:bipartitegraphs}
\qq

We introduce a graphical representation that
captures the dependencies between the measurements in $Y$
and the value vector $\Theta$, represented by $\Phi$ and $P$.
Consider a feasible decomposition of $X$ into a full-rank matrix
$P \in \mathcal{P}_F(X)$ and the corresponding $\Theta$; the 
matrix $P$ defines the sparsities of the common and innovation 
components $K_C$ and $K_j$, $1 \le j \le J$, as well as the  
joint sparsity $D = K_C + \sum_{j=1}^J K_j$.
Define the following sets of vertices: ($i$) the set of {\em value vertices}
$V_V$ has elements with indices $d \in \{1,\ldots,D\}$ representing the entries
of the value vector $\Theta(d)$, and ($ii$) the set of {\em measurement vertices}
$V_M$ has elements with indices $(j,m)$ representing the measurements
$y_j(m)$, with $j\in\Lambda$ and $m \in \{1,\ldots,M_j\}$. The cardinalities
for these sets are $|V_V|=D$ and $|V_M|=\widebar{M}$, respectively.

\begin{figure*}[bt]
\begin{center}
\epsfig{file=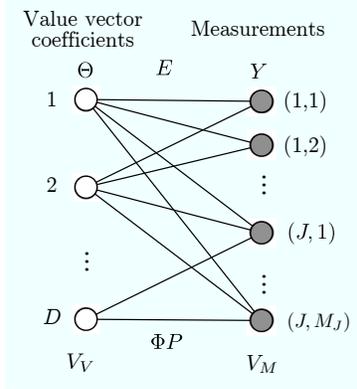,height=52mm}
\end{center}\vspace*{-4mm}
\caption{\small\sl \label{fig:bgraph}
Bipartite graph for distributed compressive sensing (DCS). The bipartite graph
$G=(V_V,V_M,E)$ indicates the relationship between the value vector coefficients
and the measurements.}
\end{figure*}

We now introduce a bipartite graph $G=(V_V,V_M,E)$, that represents the 
relationships between the entries of the value vector and the measurements 
(see \cite{DCSSensip} for details). The set of edges $E$ is defined as follows:
\begin{itemize}
\item For every $d \in \{1,2,\dots,K_C\} \subseteq V_V$ and $j \in \Lambda$ such
that column $d$ of $P_C$ does not also appear as a column of $P_j$,
we have an edge connecting $d$ to each vertex $(j,m)\in V_M$ for $1
\le m \le M_j$.
\item For every $d \in \{K_C+1, K_C+2,\dots, \numcols\} \subseteq V_V$, we
consider the sensor $j$ associated with column $d$ of $P$, and we
have an edge connecting $d$ to each vertex $(j,m) \in V_M$ for $1 \le m \le M_j$.
\end{itemize}
In words, we say that $y_j(m)$, the
$m^{th}$ measurement of sensor $j$, {\em measures} $\Theta(d)$ if the vertex
$d \in V_V$ is linked to the vertex $(j,m) \in V_M$ in the graph $G$.
An example graph for a distributed sensing setting is shown in Figure \ref{fig:bgraph}.

\qq
\subsection{Quantifying redundancies}
\label{subsec:depend}
\qq

In order to obtain sharp bounds on the number of measurements needed, our 
analysis of the measurement process must account for redundancies between the 
locations of the nonzero coefficients in the common and innovation components. 
To that end, we consider the overlaps
between common and innovation components in each signal. When we have
$z_c(n) \ne 0$ and $z_j(n) \ne 0$ for a certain signal $j$ and some index
$1\le n \le N$, we cannot recover the values of both coefficients from the
measurements of this signal alone; therefore, we will need to recover $z_c(n)$
using measurements of other signals that do not feature the same overlap.
We thus quantify the size of the overlap for all subsets of signals
$\Gamma \subset \Lambda$ under a feasible representation given by $P$ and
$\Theta$, as described in Section~\ref{sec:generalframework}.
\begin{DEFI}
The {\em overlap size} for the set of signals $\Gamma \subset \Lambda$,
denoted $K_C(\Gamma,P)$, is the number of indices in which there is overlap
between the common and the innovation component supports at all signals
$j \notin \Gamma$:
\begin{equation}
K_C(\Gamma,P) = \left|\{n \in \{1,\ldots,N\} : z_C(n) \ne 0
\textrm{ and } \forall~j \notin \Gamma,z_j(n) \ne 0\}\right|.
\label{eq:kintersect}
\end{equation}
We also define $K_C(\Lambda,P) = K_C(P)$ and $K_C(\emptyset,P) = 0$.
\end{DEFI}
For $\Gamma \subset \Lambda$, $K_C(\Gamma,P)$ provides a penalty term
due to the need for recovery of common component coefficients that are
overlapped by innovations in all other signals $j \notin \Gamma$. Intuitively, for
each entry counted in $K_C(\Gamma,P)$, some sensor in $\Gamma$ must take
one measurement to account for that entry of the common component --- it is
impossible to recover such entries from measurements made by sensors outside 
of $\Gamma$. When all signals $j \in \Lambda$ are considered, it is clear that all 
of the common component coefficients must be recovered from the obtained 
measurements.

\qq
\subsection{Measurement bounds}
\label{subsec:result}
\qq

Converse and achievable bounds for the number of measurements necessary
for DCS recovery are given below.
Our bounds consider each subset of sensors $\Gamma \subseteq \Lambda$,
since the cost of sensing the common component can be amortized across
sensors: it may be possible to reduce the rate at one sensor $j_1 \in \Gamma$
(up to a point), as long as other sensors in $\Gamma$ offset the rate reduction.
We quantify the reduction possible through the following definition.
\begin{DEFI}
The {\em conditional sparsity} of the set of signals $\Gamma$ is the number of entries of
the vector $\Theta$ that must be recovered by measurements $y_j$, $j \in \Gamma$:
$$K_{\mathrm{cond}}(\Gamma,P) = \left(\sum_{j \in \Gamma} K_j(P)\right) + K_C(\Gamma,P).$$
\end{DEFI}
The joint sparsity gives the number of degrees of freedom for the signals in
$\Lambda$, while the conditional sparsity gives the number of degrees of
freedom for signals in $\Gamma$ when the signals in $\Lambda \setminus
\Gamma$ are available as side information.
Note also that Definition~\ref{def:joint_spars} for joint sparsity
can be extended to a subset of signals $\Gamma$ by considering the number of
entries of $\Theta$ that affect these signals:
$$K_{\mathrm{joint}}(\Gamma,P) = D-K_{\mathrm{cond}}(\Lambda-\Gamma,P) = \left(\sum_{j \in \Gamma} K_j(P)\right) +K_C(P)-K_C(\Lambda\setminus\Gamma,P).$$
Note that $K_{\mathrm{cond}}(\Lambda,P) = K_{\mathrm{joint}}(\Lambda,P) = D$.

The bipartite graph introduced in Section~\ref{sec:bipartitegraphs} is the
cornerstone of Theorems~\ref{theo:achstep1}, \ref{theo:dwtb_tight_achievable_13},
and \ref{theo:cnv}, which consider whether a perfect matching can be found in
the graph; see the proofs in Appendices~\ref{app:achstep1},~\ref{app:achstep2},
and~\ref{app:cnv}, respectively, for detail.
\begin{THEO}
{\em (Achievable, known $P$)} Assume that a signal ensemble $X$ is obtained
from a common/innovation component JSM $\mathcal{P}$. Let $\mathcal{M} = (M_1,M_2,\ldots,M_J)$ be a measurement tuple, let $\{\Phi_j\}_{j \in \Lambda}$
be random matrices having $M_j$ rows of i.i.d.\
Gaussian entries for each $j \in \Lambda$, and write $Y = \Phi X$.
Suppose there exists a full rank location matrix $P \in
\mathcal{P}_F(X)$ such that
\begin{equation}
\sum_{j \in \Gamma} M_j \ge K_{\mathrm{cond}}(\Gamma,P)
\label{eq:achCondition1}
\end{equation}
for all $\Gamma \subseteq \Lambda$.
Then with probability one over $\{\Phi_j\}_{j \in \Gamma}$, there exists a unique 
solution $\widehat{\Theta}$ to the system of equations $Y = \Phi P
\widehat{\Theta}$; hence, the signal ensemble $X$ can be uniquely
recovered as $X = P \widehat{\Theta}$. \label{theo:achstep1}
\end{THEO}
\begin{THEO}
{\em (Achievable, unknown $P$)} Assume that a signal ensemble $X$ and
measurement matrices $\{\Phi_j\}_{j \in \Lambda}$ follow the assumptions of
Theorem~\ref{theo:achstep1}.
Suppose there exists a full rank
location matrix $P^* \in \mathcal{P}_F(X)$ such that
\pp
\begin{equation}
\sum_{j \in \Gamma} M_j \ge K_{\mathrm{cond}}(\Gamma,P^*)+|\Gamma|
\label{eq:dwtb_loose_achievable_condition_13}
\end{equation}
for all $\Gamma \subseteq \Lambda$. Then $X$ can be uniquely
recovered from $Y$ with probability one over $\{\Phi_j\}_{j \in \Gamma}$.\\
\label{theo:dwtb_tight_achievable_13}
\end{THEO}
\begin{THEO}
{\em (Converse)} Assume that a signal ensemble $X$ and
measurement matrices $\{\Phi_j\}_{j \in \Lambda}$ follow the assumptions of
Theorem~\ref{theo:achstep1}.
Suppose there exists a full rank location matrix $P \in
\mathcal{P}_F(X)$ such that
\begin{equation}
\sum_{j \in \Gamma} M_j < K_{\mathrm{cond}}(\Gamma,P)
\label{eq:cnvCondition1}
\end{equation}
for some $\Gamma \subseteq \Lambda$.
Then there exists a solution $\widehat{\Theta}$ such that $Y = \Phi
P \widehat{\Theta}$ but $\widehat{X} := P \widehat{\Theta} \neq X$.
\label{theo:cnv}
\end{THEO}
The identification of a feasible location matrix $P$ causes the one measurement
per sensor gap that prevents 
(\ref{eq:dwtb_loose_achievable_condition_13})--(\ref{eq:cnvCondition1})
from being a tight converse and achievable bound pair.
We note in passing that the signal recovery procedure used in
Theorem~\ref{theo:dwtb_tight_achievable_13} is akin to $\ell_0$-norm
minimization on $X$; see Appendix~\ref{app:achstep2} for details.

\subsection{Discussion}
The bounds in Theorems~\ref{theo:achstep1}--\ref{theo:cnv} are
dependent on the dimensionality of the subspaces in which the signals
reside. The number of noiseless measurements required for ensemble 
recovery is determined by the dimensionality $\dim(\mathcal{S})$ of the 
subspace $\mathcal{S}$ in the relevant
signal model, because dimensionality and sparsity play a volumetric
role akin to the entropy $H$ used to characterize rates in source coding.
Whereas in source coding each bit resolves between two options,
and $2^{NH}$ typical inputs are described using $NH$ bits~\cite{CoverThomas}, 
in CS we have $M=\dim(\mathcal{S})+O(1)$.
Similar to Slepian-Wolf coding~\cite{SW73}, the number of measurements
required for each sensor must account for the minimal features unique to that
sensor, while at the same time features that appear among multiple sensors
must be amortized over the group.

Theorems~\ref{theo:achstep1}--\ref{theo:cnv} can also be applied to the single
sensor and joint measurement settings.
In the single-signal setting (Theorem~\ref{theo:kplusone}), we will have
$x = P\theta$ with $\theta \in \real^K$, and $\Lambda = \{1\}$;
Theorem~\ref{theo:dwtb_tight_achievable_13} provides the requirement
$M \ge K+1$.
It is easy to show that the joint measurement is equivalent to the single-signal 
setting: we stack all the individual signals into a single signal vector, and in both
cases all measurements are dependent on all the entries of the signal vector. 
However, the distribution of the measurements among the available sensors is 
irrelevant in a joint measurement setting. Therefore, we only obtain a necessary 
condition $\sum_j M_j \geq D+1$ on the total number of measurements required.

\qq
\section{Practical Recovery Algorithms and Experiments} \qq
\label{sec-algs}

Although we have provided a unifying theoretical treatment for the three JSM
models, the nuances warrant further study. In particular, while 
Theorem~\ref{theo:dwtb_tight_achievable_13} highlights the basic tradeoffs 
that must be made in partitioning the measurement budget among sensors, 
the result does not by design provide insight into tractable algorithms for 
signal recovery. We believe there is additional insight to be gained by 
considering each model in turn, and while the presentation may be less 
unified, we attribute this to the fundamental diversity of problems that can 
arise under the umbrella of jointly sparse signal representations. In this 
section, we focus on tractable recovery algorithms for each model and, 
when possible, analyze the corresponding measurement requirements.

\input{jsm1section}

\qq
\subsection{Recovery strategies for common sparse supports (JSM-2)}
\label{sec-mrr2} \qq

Under the JSM-2 signal ensemble model from Section \ref{sec-ds2},
separate recovery of each signal via $\ell_0$-norm minimization would
require $\K+1$ measurements per signal, while separate recovery via
$\ell_1$-norm minimization would require $c\K$ measurements per signal.
When Theorems~\ref{theo:dwtb_tight_achievable_13} and~\ref{theo:cnv}
are applied in the context of JSM-2, the bounds for joint recovery match those
of individual recovery using $\ell_0$-norm minimization.
Within this context, it is also possible to recover one of the signals using $K+1$ measurements from the corresponding sensor, and then with the prior knowledge
of the support set $\Omega$, recover all other signals from $K$ measurements
per sensor; thus providing an additional savings of $J-1$ measurements~\cite{Rob}.
Surprisingly, we will demonstrate below that for large $J$, the common support set can actually be recovered using only one measurement per sensor and algorithms that are computationally tractable.

The algorithms we propose are inspired by conventional greedy pursuit
algorithms for CS (such as OMP~\cite{TroppOMP}). In the single-signal
case, OMP iteratively constructs the sparse support set $\Omega$;
decisions are based on inner products between the columns of
$\Phi$ and a residual. In the multi-signal case, there are more
clues available for determining the elements of $\Omega$.

\qq
\subsubsection{Recovery via Trivial Pursuit (TP)} \qq

When there are many correlated signals in the ensemble, a simple
non-iterative greedy algorithm based on inner products will
suffice to recover the signals jointly. For simplicity but
without loss of generality, we assume that an equal number of
measurements $\M_\j = \M$ are taken of each signal. We write
$\Phi_\j$ in terms of its columns, with $\Phi_\j = \left[
\phi_{\j,1}, \phi_{\j,2}, \dots, \phi_{\j,\N} \right]$.

\bigskip\centerline{\bf Trivial Pursuit (TP) Algorithm for JSM-2}\vspace*{-3mm}

\begin{enumerate}
\item{\bf Get greedy:}  Given all of the measurements,
compute the test statistics
\vspace{-2mm}
\begin{equation}
\xi_\n = \frac{1}{\J}\sum_{\j=1}^\J \langle y_\j ,  \phi_{\j,\n} \rangle^2,
\qquad \n \in \{1,2,\dots,\N\},
\label{eq:oneshot}
\end{equation}
\vspace{-2mm}
and estimate the elements of the common coefficient support set by
\[
\widehat{\Omega} = \{ \n ~\mbox{having one of the $\K$ largest}~ \xi_\n
\}.
\]
\vspace{-2mm}
\end{enumerate}

When the sparse, nonzero coefficients are sufficiently generic (as
defined below), we have the following surprising result, which is
proved in Appendix~\ref{ap:onepass}.

\begin{THEO}
\label{theo:onepass} Let $\Psi$ be an orthonormal basis for
$\mathbb{R}^\N$, let the measurement matrices $\Phi_\j$ contain
i.i.d.\ Gaussian entries, and assume that the nonzero coefficients
in the $\theta_\j$ are i.i.d.\ Gaussian random variables. Then
with $\M \ge 1$ measurements per signal, TP recovers $\Omega$
with probability approaching one as $\J \rightarrow \infty$.
\end{THEO}

In words, with {\em fewer} than $\K$ measurements per sensor, it is
actually possible to recover the sparse support set $\Omega$ under
the JSM-2 model.\footnote{ One can also show the somewhat stronger
result that, as long as $\sum_j M_j \gg N$, TP recovers $\Omega$
with probability approaching one. We have omitted this additional
result for brevity.} Of course, this approach does not recover the
$\K$ coefficient values for each signal; at least $\K$ measurements
per sensor are required for this.

\begin{COROLLARY}
\label{cor:onepass2} Assume that the nonzero coefficients in the
$\theta_\j$ are i.i.d.\ Gaussian random variables. Then the following
statements hold:
\begin{enumerate}
\item Let the measurement matrices $\Phi_\j$ contain i.i.d.\
Gaussian entries, with each matrix having an overmeasuring factor of
$c=1$ (that is, $\M_\j=\K$ for each measurement matrix $\Phi_\j$).
Then TP recovers all signals from the ensemble $\{x_\j\}$ with
probability approaching one as $\J \rightarrow \infty$.

\item Let $\Phi_\j$ be a measurement matrix with overmeasuring
factor $c < 1$ (that is, $\M_\j < \K$), for some $\j \in \Lambda$. Then with
probability one, the signal $x_\j$
cannot be uniquely recovered by any algorithm for any value of $\J$.
\end{enumerate}
\end{COROLLARY}

The first statement is an immediate corollary of
Theorem~\ref{theo:onepass}; the second statement follows because
each equation $y_\j = \Phi_\j x_\j$ would be underdetermined even if
the nonzero indices were known. Thus, under the JSM-2 model, the
TP algorithm asymptotically performs as well as an
oracle decoder that has prior knowledge of the locations of the
sparse coefficients. From an information theoretic perspective,
Corollary~\ref{cor:onepass2} provides tight achievable and converse
bounds for JSM-2 signals. We should note that the
theorems in this section have a slightly different flavor than
Theorem~\ref{theo:dwtb_tight_achievable_13} and~\ref{theo:cnv}, which
ensure recovery of {\em any} sparse signal ensemble, given a suitable
set of measurement matrices.
Theorem~\ref{theo:onepass} and Corollary~\ref{cor:onepass2} above, in
contrast, rely on a random signal model and do not guarantee
simultaneous performance for all sparse signals under any particular
measurement ensemble. Nonetheless, we feel this result is worth
presenting to highlight the strong subspace concentration behavior
that enables the correct identification of the common support.

In the technical reports~\cite{DCSTR06,CLTTR2005}, we derive an
approximate formula for the probability of error in recovering the common
support set $\Omega$ given $\J$, $\K$, $\M$, and $\N$.
While theoretically interesting and potentially practically useful, these results
require $\J$ to be large. Our numerical experiments show that the number
of measurements required for recovery using TP decreases quickly as $\J$ 
increases. However, in the case of small $\J$, TP performs poorly. 
Hence, we propose next an alternative recovery technique based on 
simultaneous greedy pursuit that performs well for small $\J$.

\qq
\subsubsection{Recovery via iterative greedy pursuit}
\label{sec:somp-cs} \qq

In practice, the common sparse support among the $\J$ signals
enables a fast iterative algorithm to recover all of the signals
jointly.  Tropp and Gilbert have proposed one such algorithm, called
{\em Simultaneous Orthogonal Matching Pursuit} (SOMP)
\cite{Tropp05}, which can be readily applied in our DCS framework.
SOMP is a variant of OMP that seeks to identify $\Omega$ one element
at a time. A similar simultaneous sparse approximation algorithm
has been proposed using convex optimization
\cite{TroppSimConvex}. We dub the DCS-tailored SOMP
algorithm DCS-SOMP.

To adapt the original SOMP algorithm to our setting, we first extend
it to cover a different measurement matrix $\Phi_\j$ for each signal
$x_\j$. Then, in each DCS-SOMP iteration, we select the column index
$\n \in \{1,2,\dots,\N\}$ that accounts for the greatest amount of
residual energy across {\em all} signals. As in SOMP, we
orthogonalize the remaining columns (in each measurement matrix)
after each step; after convergence we obtain an expansion of the
measurement vector $y_j$ on an orthogonalized subset of the columns of
basis vectors. To obtain the expansion coefficients in the sparse
basis, we then reverse the orthogonalization process using the QR
matrix factorization. Finally, we again assume that
$\M_\j = \M$ measurements per signal are taken.

\vspace{5mm}
\bigskip\centerline{\bf DCS-SOMP Algorithm for JSM-2}\vspace*{-3mm}
\begin{enumerate}
\item{\bf Initialize:} Set the iteration counter $\ell = 1$. For
each signal index $\j \in \Lambda$, initialize the
orthogonalized coefficient vectors $\widehat{\beta_\j} = 0$,
$\widehat{\beta}_\j \in \mathbb{R}^\M$; also initialize the set of
selected indices $\hatomega = \emptyset$. Let $r_{\j,\ell}$ denote
the residual of the measurement $y_\j$ remaining after the first
$\ell$ iterations, and initialize $r_{\j,0} = y_\j$.
\item{\bf Select} the dictionary vector that maximizes the value
of the sum of the magnitudes of the projections of the residual, and
add its index to the set of selected indices
\vspace{-2mm}
\begin{eqnarray*}
\n_\ell &=& \arg\!\!\!\!\!\max_{\n \in \{1,\ldots,N\}}\sum_{\j=1}^\J
\frac{|\langle r_{\j,\ell-1},\phi_{\j,\n} \rangle|}{\|\phi_{\j,\n}\|_2}, \\
\hatomega &=& [\hatomega~~\n_\ell].
\vspace{-2mm}
\end{eqnarray*}
\item{\bf Orthogonalize} the selected basis vector against the
orthogonalized set of previously selected dictionary vectors
\vspace{-2mm}
\begin{equation*}
\gamma_{\j,\ell} = \phi_{\j,\n_\ell} -
\sum_{t=0}^{\ell-1}\frac{\langle
\phi_{\j,\n_\ell},\gamma_{\j,t}\rangle}{\|\gamma_{\j,t}\|_2^2}\gamma_{\j,t}.
\vspace{-1mm}
\end{equation*}
\item{\bf Iterate:} Update the estimate of the coefficients for
the selected vector and residuals
\vspace{-1mm}
\begin{eqnarray*}
\widehat{\beta}_{\j}(\ell) &=& \frac{\langle
r_{\j,\ell-1},\gamma_{\j,\ell}\rangle}{\|\gamma_{\j,\ell}\|_2^2}, \\
r_{\j,\ell} &=& r_{\j,\ell-1}-\frac{\langle
r_{\j,\ell-1},\gamma_{\j,\ell}\rangle}
{\|\gamma_{\j,\ell}\|_2^2}\gamma_{\j,\ell}.
\vspace{-2mm}
\end{eqnarray*}
\item {\bf Check for convergence:} If $\|r_{\j,\ell}\|_2 >
\epsilon \|{y_\j}\|_2$ for all $\j$, then increment $\ell$ and go
to Step 2; otherwise, continue to Step 6. The parameter $\epsilon$
determines the target error power level allowed for algorithm
convergence.
\item {\bf De-orthogonalize:} Consider the relationship between
$\Gamma_\j = [\gamma_{\j,1},\gamma_{\j,2},\ldots,\gamma_{\j,\M}]$
and the $\Phi_\j$ given by the QR factorization
$\Phi_{\j,\hatomega} = \Gamma_\j R_\j$,
where $\Phi_{\j,\hatomega} =
[\phi_{\j,\n_1},\phi_{\j,\n_2},\ldots,\phi_{\j,\n_\M}]$ is the
so-called {\em mutilated basis}.\footnote{We define a {\em mutilated
basis} $\Phi_\Omega$ as a subset of the basis vectors from $\Phi =
[\phi_1,\phi_2,\ldots,\phi_\N]$ corresponding to the indices given
by the set $\Omega = \{\n_1,\n_2,\ldots, \n_\M\}$, that is,
$\Phi_\Omega = [\phi_{\n_1},\phi_{\n_2},\ldots,\phi_{\n_\M}]$. This
concept can be extended to vectors in the same manner.} Since $y_\j
= \Gamma_\j\beta_\j = \Phi_{\j,\hatomega}x_{\j,\hatomega} =
\Gamma_\j R_\j x_{\j,\hatomega}$, where $x_{\j,\hatomega}$ is the
mutilated coefficient vector, we can compute the signal estimates
$\{\widehat{x}_\j\}$ as
\vspace{-2mm}
\begin{equation*}
\widehat{x}_{\j,\hatomega} = R_\j^{-1}\widehat{\beta}_\j, \\
\end{equation*}
\vspace{-2mm}
where $\widehat{x}_{\j,\hatomega}$ is the mutilated version of
the sparse coefficient vector $\widehat{x}_\j$.
\end{enumerate}
In practice, we obtain $\widehat{c}\K$ measurements from each signal 
$x_\j$ for some value of $\widehat{c}$.  We then use DCS-SOMP to 
recover the $\J$ signals jointly.
We orthogonalize because as the number of iterations approaches $\M$
the norms of the residues of an orthogonal pursuit decrease faster
than for a non-orthogonal pursuit; indeed, due to Step 3 the algorithm can
only run for up to $M$ iterations. The computational complexity of this
algorithm is $O(JNM^2)$, which matches that of separate
recovery for each signal  while reducing the
required number of measurements.

Thanks to the common sparsity structure among the signals, we
believe (but have not proved) that DCS-SOMP will succeed with
$\widehat{c} < c(S)$.
Empirically, we have observed that a small number of measurements
proportional to $\K$ suffices for a moderate number of sensors
$J$. Based on our observations, described in 
Section~\ref{subsec:jsm2examples}, we conjecture that $\K+1$ 
measurements per sensor suffice as $\J \rightarrow \infty$. Thus, this 
efficient greedy algorithm enables an overmeasuring factor
$\widehat{c}=(\K+1)/\K$ that approaches $1$ as $\J$, $\K$, and
$\N$ increase.

\qq
\subsubsection{Simulations for JSM-2} \label{subsec:jsm2examples} \qq

\begin{figure*}[t]
\begin{center}
\epsfig{file=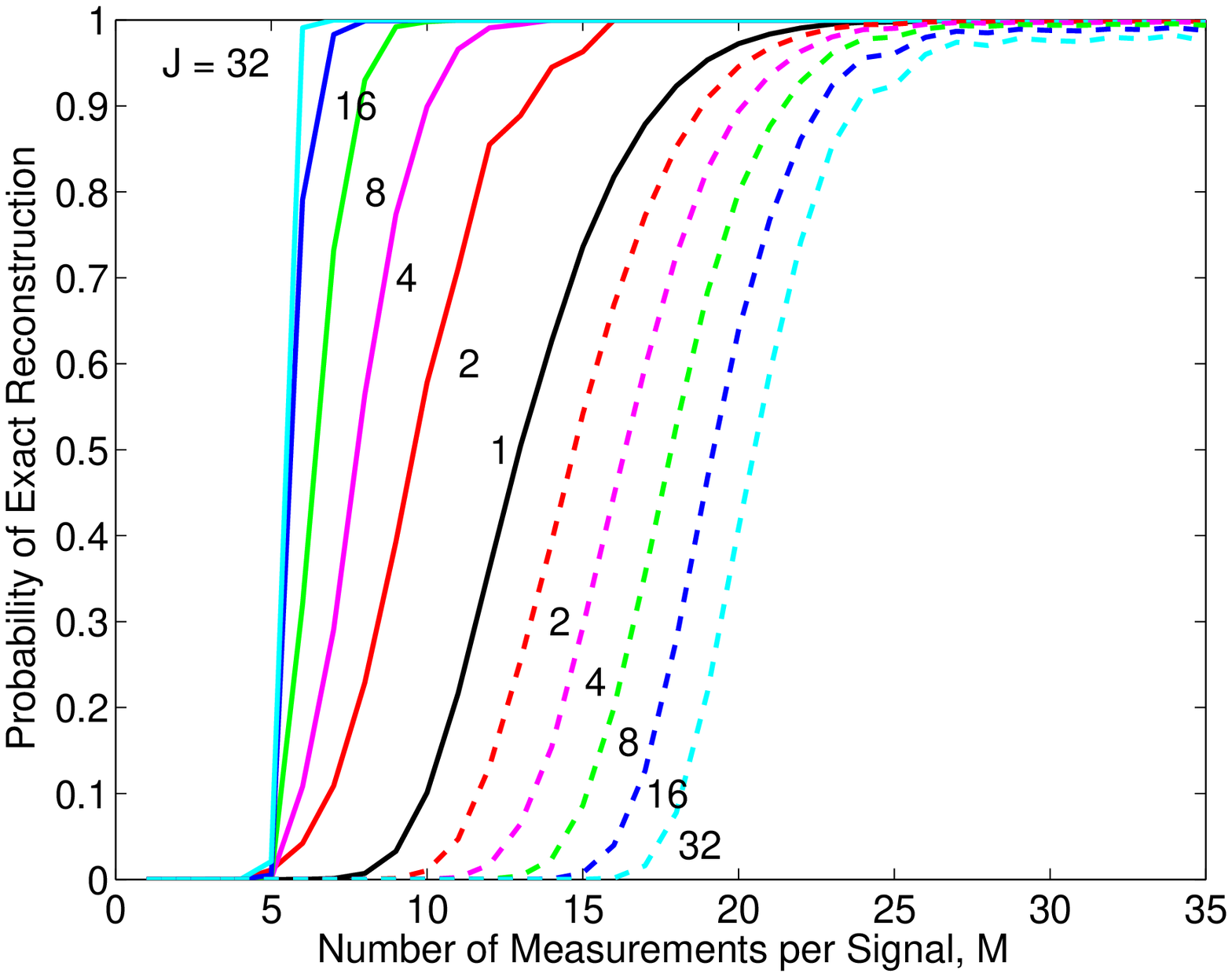,height=60mm}
\end{center}\vspace*{-7mm}
\caption{\small\sl \label{fig-somp} Recovering a signal ensemble with
common sparse supports (JSM-2).  We plot the probability of perfect
recovery via DCS-SOMP (solid lines) and separate CS
recovery (dashed lines) as a function of the number of
measurements per signal $\M$ and the number of signals $\J$. We fix
the signal length to $\N = 50$, the sparsity to $\K = 5$, and
average over 1000 simulation runs. An oracle encoder that knows the
positions of the large signal expansion coefficients would use 5
measurements per signal.}
\end{figure*}

We now present simulations comparing separate CS recovery
versus joint DCS-SOMP recovery for a JSM-2 signal ensemble.
Figure~\ref{fig-somp} plots the probability of perfect
recovery corresponding to various numbers of measurements $\M$
as the number of sensors varies from $\J = 1$ to $32$, over 1000
trials in each case.  We fix the signal lengths at $\N = 50$ and the
sparsity of each signal to $\K = 5$.

With DCS-SOMP, for perfect recovery of all signals the
average number of measurements per signal decreases as a function
of $\J$. The trend suggests that for large $\J$ close to
$\K$ measurements per signal should suffice. On the contrary, with
separate CS recovery, for perfect recovery of all
signals the number of measurements per sensor {\em increases} as a
function of $\J$. This occurs because each
signal experiences an independent probability $p \le 1$ of
successful recovery; therefore the overall probability of
complete success is $p^\J$. Consequently, each sensor must
compensate by making additional measurements. This phenomenon
further motivates joint recovery under JSM-2.

Finally, we note that we can use algorithms other than DCS-SOMP to recover
the signals under the JSM-2 model. Cotter et al.~\cite{Cotter05} have proposed
additional algorithms (such as M-FOCUSS) that iteratively
eliminate basis vectors from the dictionary and converge to the set of
sparse basis vectors over which the signals are supported.
We hope to extend such algorithms to JSM-2 in future work.

\qq
\subsection{Recovery strategies for nonsparse common component \\
+ sparse innovations (JSM-3)}
\label{sec-mrr3} \qq

The JSM-3 signal ensemble model from Section \ref{sec-ds3} provides
a particularly compelling motivation for joint recovery. Under this
model, no individual signal $x_\j$ is sparse, and so recovery of
each signal separately would require fully $\N$ measurements per
signal. As in the other JSMs, however, the commonality among the
signals makes it possible to substantially reduce this number.
Again, the potential for this savings is evidenced by specializing
Theorem~\ref{theo:dwtb_tight_achievable_13} to the context of JSM-3.

\begin{COROLLARY}
If $\Phi_j$ is a random Gaussian matrix for all $j \in \Lambda$,
$\mathcal{P}$ is defined by JSM-3, and
\begin{eqnarray}
\sum_{\j\in\Gamma}\M_\j &\ge& \sum_{\j\in\Gamma}\K_\j + K_C(\Gamma,P)
+ |\Gamma|, ~~ \Gamma \in \Lambda, \label{eq:jsm3th0-1} \\
\sum_{\j\in\Lambda}\M_\j &\ge& \sum_{\j\in\Lambda}\K_\j + N + J - K_R, \label{eq:jsm3th0-2}
\end{eqnarray}
then the signal ensemble $X$ can be uniquely recovered from $Y$ with
probability one.
\label{th:jsm3th0}
\end{COROLLARY}

This suggests that the number of measurements of an individual signal 
can be substantially decreased, as long as the total number of measurements 
is sufficiently large to capture enough
information about the nonsparse common component $z_C$. The term
$K_R$ denotes the number of indices where the common and all innovation 
components overlap, and appears due to the sparsity reduction that can be 
performed at the common component before recovery. We also note that when 
the supports of the innovations are independent, as $J \to \infty$, it becomes
increasingly unlikely that a given index will be included in all innovations,
and thus the terms $K_C(\Gamma,P)$ and $K_R$ will go to zero. On the
other hand, when the supports are completely matched (implying
$K_j = K$, $j \in \Lambda$), we will have $K_R = K$, and after sparsity
reduction has been addressed, $K_C(\Gamma,P) = 0$ for all
$\Gamma \subseteq \Lambda$.

\qq
\subsubsection{Recovery via Transpose Estimation of Common Component
(TECC)} \label{subsec:nonsparse} \qq

Successful recovery of the signal ensemble $\{x_\j\}$ requires
recovery of both the nonsparse common component $z_C$ and the sparse
innovations $\{z_\j\}$. To help build intuition about how we might
accomplish signal recovery using far fewer than $\N$ measurements
per sensor, consider
the following thought experiment.

If $z_C$ were known, then each innovation $z_\j$ could be estimated
using the standard single-signal CS machinery on the adjusted
measurements
$y_\j - \Phi_\j z_C = \Phi_\j z_\j$.
While $z_C$ is not known in advance, it can be {\em estimated} from
the measurements. In fact, across all $\J$ sensors, a total of
$\sum_{\j\in\Lambda} \M_\j$ random projections of $z_C$ are observed (each
corrupted by a contribution from one of the $z_\j$). Since $z_C$ is
not sparse, it cannot be recovered via CS techniques, but when the
number of measurements is sufficiently large ($\sum_{\j\in\Lambda} \M_\j \gg
\N$), $z_C$ can be estimated using standard tools from linear
algebra.  A key requirement for such a method to succeed in
recovering $z_C$ is that each $\Phi_\j$ be different, so that their
rows combine to span all of $\mathbb{R}^\N$. In the limit (again,
assuming the sparse innovation coefficients are well-behaved), the
common component $z_C$ can be recovered while still allowing each
sensor to operate at the minimum measurement rate dictated by the
$\{z_\j\}$. A prototype algorithm is listed below, where we assume
that each measurement matrix $\Phi_\j$ has i.i.d.\
$\mathcal{N}(0,\sigma_\j^2)$ entries.

\bigskip\centerline{\bf TECC Algorithm for JSM-3}\vspace*{-3mm}
\begin{enumerate}

\item{\bf Estimate common component:} Define the matrix
$\widehat{\Phi}$ as the vertical concatenation of the regularized
individual measurement matrices $\widehat{\Phi}_\j =
\frac{1}{\M_\j\sigma_\j^2}\Phi_\j$, that is, $\widehat{\Phi} =
[\widehat{\Phi}_1\trans,\widehat{\Phi}_2\trans,\ldots,\widehat{\Phi}_\J\trans]\trans$.
Calculate the estimate of the common component as $\widehat{z}_C =
\frac{1}{\J} \widehat{\Phi}\trans Y$.

\item{\bf Estimate measurements generated by innovations:} Using
the previous estimate, subtract the contribution of the common
part from the measurements and generate estimates for the
measurements caused by the innovations for each signal:
$\widehat{y}_\j = y_\j - \Phi_\j \widehat{z}_C$.

\item{\bf Recover innovations:} Using a standard single-signal CS
recovery algorithm,\footnote{For tractable analysis of the TECC algorithm,
the proof of Theorem~\ref{theo:nonsparse} employs a least-squares variant of
$\ell_0$-norm minimization.} 
obtain estimates of the innovations
$\widehat{z}_\j$ from the estimated innovation measurements
$\widehat{y}_\j$.

\item{\bf Obtain signal estimates:} Estimate each signal as the
sum of the common and innovations estimates; that is,
$\widehat{x}_\j = \widehat{z}_C + \widehat{z}_\j$.
\end{enumerate}

The following theorem, proved in Appendix \ref{ap:jsm3}, shows
that asymptotically, by using the TECC algorithm, each sensor needs
to only measure at the rate dictated by the sparsity $\K_\j$.

\begin{THEO} \label{theo:nonsparse}
Assume that the nonzero expansion coefficients of the sparse
innovations $z_\j$ are i.i.d.\ Gaussian random variables and that
their locations are uniformly distributed on $\{1,2,\ldots,\N\}$. Let
the measurement matrices $\Phi_\j$ contain i.i.d.\
$\mathcal{N}(0,\sigma_\j^2)$ entries with $\M_\j \ge \K_\j+1$. Then
each signal $x_\j$ can be recovered using the TECC algorithm with
probability approaching one as $\J \rightarrow \infty$.
\end{THEO}

For large $\J$, the measurement
rates permitted by Theorem~\ref{theo:nonsparse}
are the best possible for {\em any} recovery strategy for
JSM-3 signals, even neglecting the presence of the nonsparse
component. These rates meet the minimum bounds suggested
by Corollary~\ref{th:jsm3th0}, although again Theorem~\ref{theo:nonsparse}
is of a slightly different flavor, as it does not provide a uniform
guarantee for all sparse signal ensembles under any particular
measurement matrix collection.
The CS technique employed in Theorem~\ref{theo:nonsparse} involves
combinatorial searches that estimate the innovation components; we
have provided the theorem simply as support for our intuitive
development of the TECC algorithm. More efficient techniques could
also be employed (including several proposed for CS in the presence
of noise~\cite{DonohoECS,CandesCS,HauptNowak}). It is
reasonable to expect similar behavior; as the error in estimating
the common component diminishes, these techniques should perform
similarly to their noiseless analogues.

\qq
\subsubsection{Recovery via Alternating Common and Innovation
Estimation (ACIE)} \qq

The preceding analysis
demonstrates that the number of required measurements in JSM-3 can
be substantially reduced through joint recovery. While
Theorem~\ref{theo:nonsparse} shows theoretical gains as $\J
\rightarrow \infty$, practical gains can also be realized with a
moderate number of sensors.
In particular, suppose in the TECC algorithm that the initial estimate
$\widehat{z}_C$ is not accurate enough to enable correct identification
of the sparse innovation supports $\{\Omega_\j\}$. In such a case, it may still be
possible for a rough approximation of the innovations $\{z_\j\}$ to
help refine the estimate $\widehat{z}_C$.  This in turn could help
to refine the estimates of the innovations.

The Alternating Common and Innovation Estimation (ACIE) algorithm
exploits the observation that once the basis vectors comprising
the innovation $z_\j$ have been identified in the index set
$\Omega_\j$, their effect on the measurements $y_\j$ can be
removed to aid in estimating $z_C$.  Suppose that we have an
estimate for these innovation basis vectors in $\hatomega_\j$.  We
can then partition the measurements into two parts: the projection
into span$(\{\phi_{\j,\n}\}_{\n \in \hatomega_\j})$ and the
component orthogonal to that span. We build a basis for the
$\mathbb{R}^{\M_\j}$ where $y_\j$ lives:
\begin{equation*}
B_\j = [\Phi_{\j,\hatomega_\j}~Q_\j],
\end{equation*} where $\Phi_{\j,\hatomega_\j}$ is the
mutilated matrix $\Phi_\j$ corresponding to the indices in
$\hatomega_\j$, and the $\M_\j \times (\M_\j-|\hatomega_\j|)$ matrix $Q_\j$
has orthonormal columns that span the orthogonal complement of
$\Phi_{\j,\hatomega_\j}$.

This construction allows us to remove the projection of the
measurements into the aforementioned span to obtain measurements
caused exclusively by vectors not in $\hatomega_\j$:
\begin{equation}
\widetilde{y}_\j = Q_\j\trans y_\j~\textrm{and}~
\widetilde{\Phi}_\j = Q_\j\trans \Phi_\j. \label{eq:mody}
\end{equation}
These modifications enable the sparse decomposition of the
measurement, which now lives in $\mathbb{R}^{\M_\j-|\hatomega_\j|}$, to
remain unchanged:
\vspace{-2mm}
\begin{equation*}
\widetilde{y_\j} = \sum_{\n=1}^\N\alpha_\j\widetilde{\phi}_{\j,\n}.
\end{equation*}
Thus, the modified measurements $\widetilde{Y} =
\left[\widetilde{y}_1\trans ~ \widetilde{y}_2\trans ~
\ldots~\widetilde{y}_\J\trans \right]\trans$ and modified
measurement matrix $\widetilde{\Phi} =
\left[\widetilde{\Phi}_1\trans ~ \widetilde{\Phi}_2\trans ~
\ldots~\widetilde{\Phi}_\J\trans \right]\trans$ can be used to
refine the estimate of the
common component of the signal,
\vspace{-2mm}
\begin{equation}
\widetilde{z_C} = \widetilde{\Phi}^\dagger\widetilde{Y},
\label{eq:spinv}
\end{equation}
where $A^\dagger = (A\trans A)^{-1}A\trans$ denotes the
pseudoinverse of matrix $A$.

When the innovation support estimate is correct
($\hatomega_\j = \Omega_\j$), the measurements $\widetilde{y_\j}$
will describe only the common component $z_C$. If this is true for
every signal $\j$ and the number of remaining measurements
$\sum_{\j=1}^J (M_\j -\K_j) \ge \N$, then $z_C$ can be perfectly
recovered via (\ref{eq:spinv}). However, it may be difficult to
obtain correct estimates for all signal supports in the first
iteration of the algorithm, and so we find it preferable to refine
the estimate of the support by executing several iterations.

\bigskip\centerline{\bf ACIE Algorithm for JSM-3}\vspace*{-3mm}
\begin{enumerate}
\item{\bf Initialize:} Set $\hatomega_\j = \emptyset$ for each
$\j$. Set the iteration counter $\ell = 1$.

\item{\bf Estimate common component:} Update estimate
$\widetilde{z_C}$ according to (\ref{eq:mody})--(\ref{eq:spinv}).

\item{\bf Estimate innovation supports:} For each sensor $\j$,
after subtracting the contribution $\widetilde{z_C}$ from the
measurements, $\widehat{y_\j} = y_\j - \Phi_\j \widetilde{z_C}$,
estimate the support of each signal innovation
$\hatomega_\j$.

\item{\bf Iterate:} If $\ell < L$, a preset number of iterations,
then increment $\ell$ and return to Step 2. Otherwise proceed to
Step 5.

\item{\bf Estimate innovation coefficients:} For each
$\j$,
estimate the coefficients for the indices in $\hatomega_\j$,
\vspace{-2mm}
\begin{equation*}
\widehat{z}_{\j,\hatomega_\j} = \Phi_{\j,\hatomega_\j}^\dagger (y_\j -
\Phi_\j \widetilde{z_C}),
\end{equation*}
where $\widehat{z}_{\j,\hatomega_\j}$ is a mutilated version of
the innovation's sparse coefficient vector estimate $\widehat{z}_\j$.

\item {\bf Recover signals:} Compute the estimate of each
signal as $\widehat{x}_\j = \widetilde{z_C} + \widehat{z}_\j$.

\end{enumerate}

Estimation of the supports in Step 3 can be accomplished
using a variety of techniques.  We propose to run
a fixed number of iterations
of OMP; if the supports of the innovations are known to match across
signals --- as in JSM-2 --- then more powerful
algorithms like SOMP can be used. The ACIE algorithm is similar in 
spirit to other iterative estimation algorithms, such as turbo 
decoding~\cite{Berrou1993}.

\qq
\subsubsection{Simulations for JSM-3} \qq

We now present simulations of JSM-3 recovery for the
following scenario. Consider $\J$ signals of length $\N = 50$
containing a common white noise component $z_C(\n) \sim
\mathcal{N}(0,1)$ for $\n \in \{1,\ldots,N\}$.  Each innovations
component $z_\j$ has sparsity $\K=5$ (once again in the time
domain), resulting in $x_\j = z_C+z_\j$. The signals are generated
according to the model used in Section~\ref{subsec-examplesjsm1}.

We study two different cases. The first is an extension of JSM-1:
we select the supports for the various innovations separately
and then apply OMP to each signal in Step 3 of the
ACIE algorithm in order to estimate its innovations component. The
second case is an extension of JSM-2: we select one common support
for all of the innovations across the signals and then apply the
DCS-SOMP algorithm (Section \ref{sec:somp-cs}) to estimate the
innovations in Step 3. In both cases we use $L = 10$ iterations of ACIE.
We test the algorithms for different numbers of signals $\J$ and
calculate the probability of correct recovery as a function
of the (same) number of measurements per signal $\M$.

\begin{figure*}[bt]
\begin{center}
\begin{tabular}{cc}
\epsfig{file=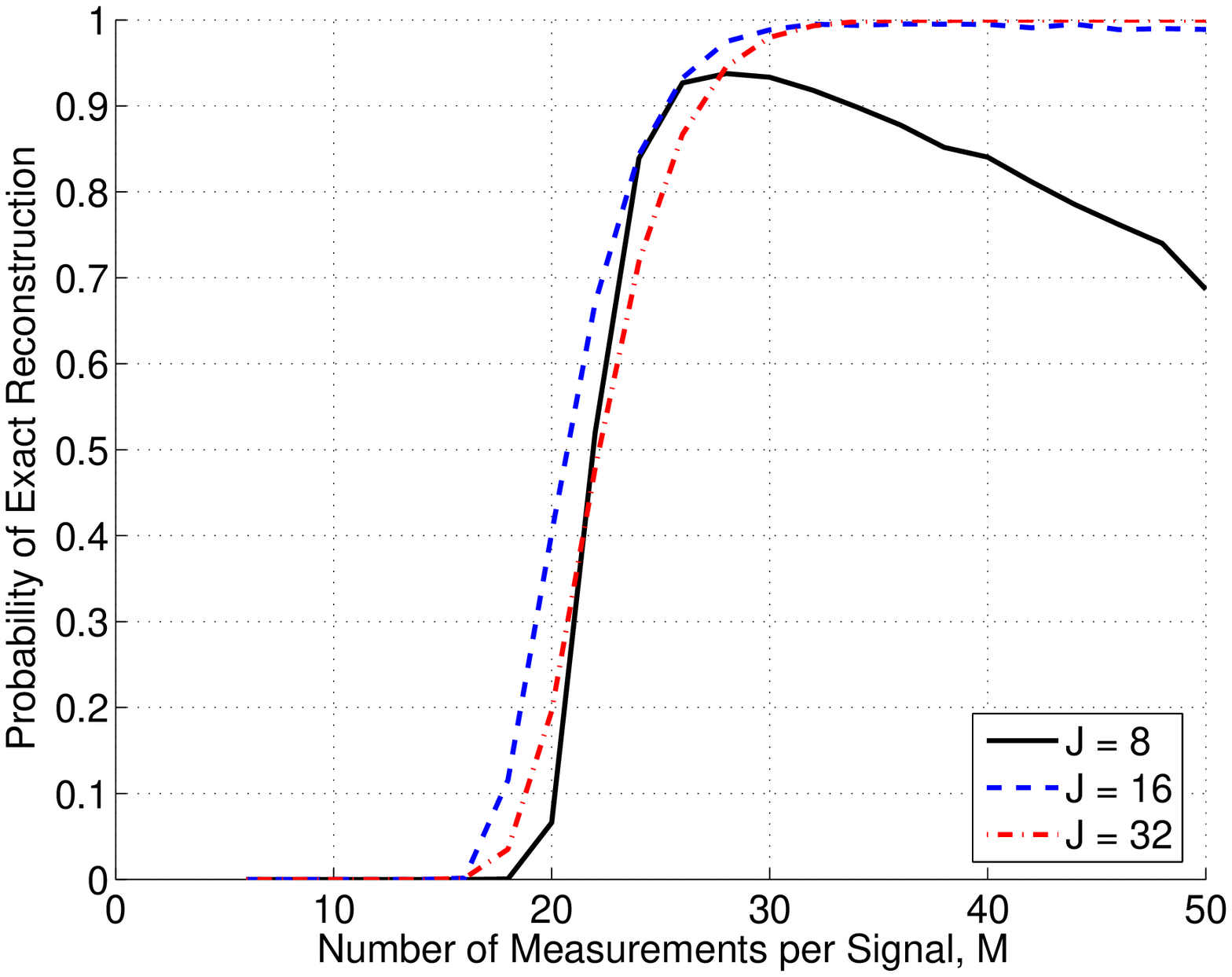,height=60mm} &
\epsfig{file=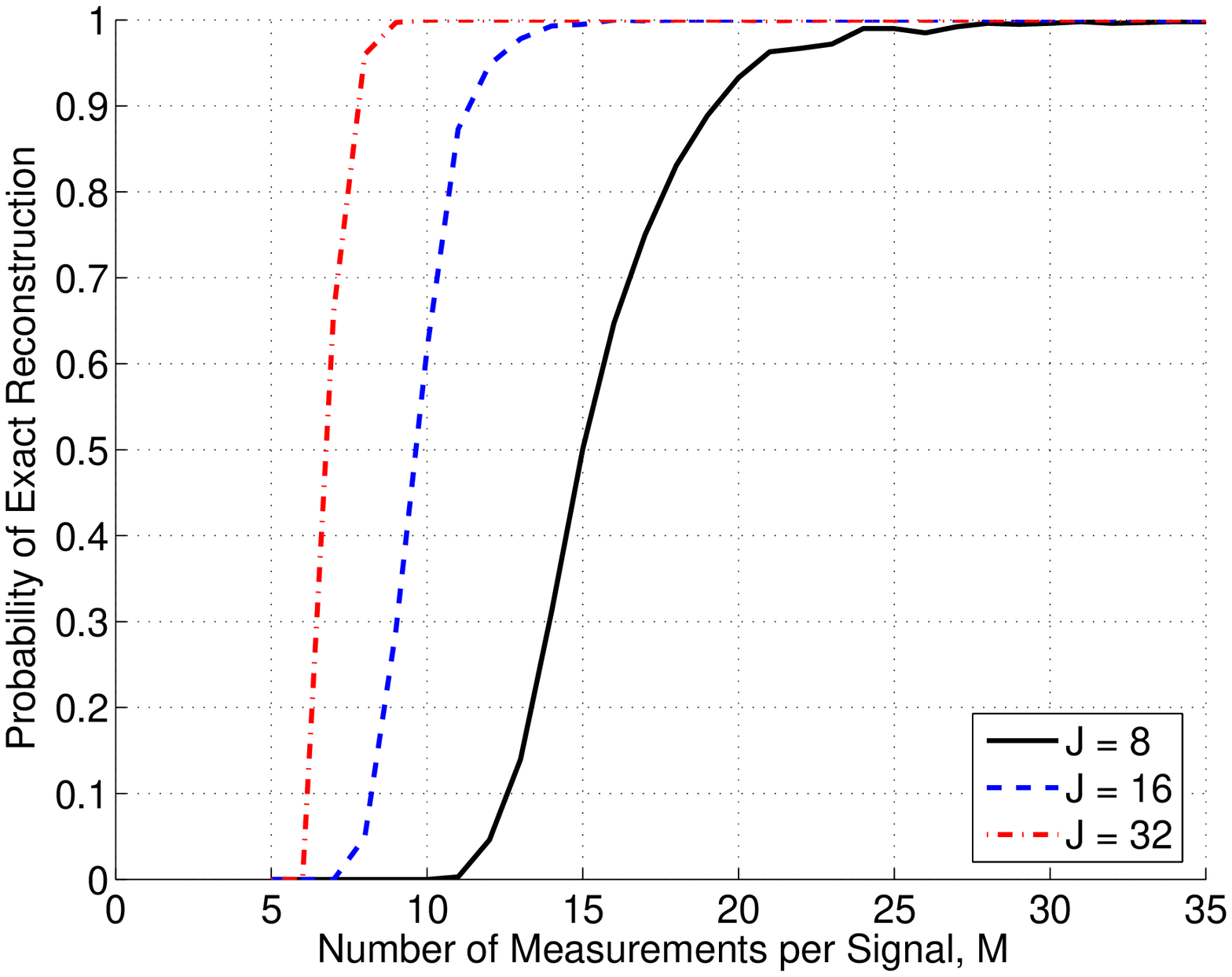,height=60mm} \\
(a) & (b) \\
\end{tabular}
\end{center}\vspace*{-7mm}
\caption{\small\sl \label{fig:jsm3}
Recovering a signal ensemble with nonsparse common component
and sparse innovations (JSM-3) using ACIE. (a) recovery using OMP
separately on each signal in Step 3 of the ACIE algorithm (innovations
have arbitrary supports). (b) recovery using DCS-SOMP jointly on all
signals in Step 3 of the ACIE algorithm (innovations have identical
supports). Signal length $\N=50$, sparsity $\K=5$. The common structure
exploited by DCS-SOMP enables dramatic savings in the number of
measurements. We average over 1000 simulation runs.}
\end{figure*}

Figure~\ref{fig:jsm3}(a) shows that, for sufficiently large $\J$, we
can recover all of the signals with significantly fewer than $\N$
measurements per signal. As $\J$ grows, it becomes more difficult to perfectly
recover all $\J$ signals. We believe this is inevitable, because
even if $z_C$ were known without error, then perfect ensemble
recovery would require the successful execution of $\J$ {\em
independent} runs of OMP. Second, for small $\J$, the probability of
success can decrease at high values of $\M$. We believe this
behavior is due to the fact that initial errors in estimating $z_C$
may tend to be somewhat sparse (since $\widehat{z}_C$ roughly
becomes an average of the signals $\{x_\j\}$), and these sparse
errors can mislead the subsequent OMP processes. For more moderate
$\M$, it seems that the errors in estimating $z_C$ (though greater)
tend to be less sparse. We expect that a more sophisticated
algorithm could alleviate such a problem; the
problem is also mitigated at higher $\J$.

Figure \ref{fig:jsm3}(b) shows that when the sparse innovations
share common supports we see an even greater savings. As a point of
reference, a traditional approach to signal acquisition would require
$1600$ total measurements to recover these $\J=32$ nonsparse
signals of length $\N=50$. Our approach requires only approximately
$10$ random measurements per sensor  --- a total of $320$
measurements --- for high probability of recovery.

\qq
\section{Discussion and Conclusions}
\label{sec-discussion}
\qq

In this paper we have extended the
theory and practice of compressive sensing to multi-signal,
distributed settings. The
number of noiseless measurements required for ensemble recovery
is determined by the dimensionality of the subspace in the relevant
signal model, because dimensionality and sparsity play a volumetric
role akin to the entropy used to characterize rates in source coding.
Our three example joint sparsity models (JSMs)
for signal ensembles with both intra- and inter-signal correlations
capture the essence of real physical scenarios, illustrate the basic
analysis and algorithmic techniques, and indicate the significant
gains to be realized from joint recovery.  In some sense, distributed
compressive sensing (DCS) is a framework for distributed compression
of sources with memory, which has remained a challenging problem for
some time.

In addition to offering substantially reduced measurement rates, the
DCS-based distributed source coding schemes we develop here share
the properties of CS mentioned in Section~\ref{sec-cs}. 
Two additional properties of DCS make it well-matched to distributed
applications such as sensor networks and arrays
\cite{estrin_2002,pottie:cacm00}. First, each sensor encodes its
measurements separately, which eliminates the need for inter-sensor
communication. Second, DCS distributes its computational
complexity asymmetrically, placing most of it in the joint decoder,
which will often have more computational resources than
any individual sensor node.  The encoders are very simple; they
merely compute incoherent projections with their signals and make no
decisions.

There are many opportunities for applications and extensions of
these ideas.
First, natural signals are not
exactly sparse but rather can be better modeled as $\ell_p$-compressible
with $0 < p \le 1$.  Roughly speaking, a signal in a {\em
weak}-$\ell_p$ ball has coefficients that decay as $n^{-1/p}$ once
sorted according to magnitude~\cite{DonohoCS}. The key concept is
that the ordering of these coefficients is important. For JSM-2, we
can extend the notion of simultaneous sparsity for $\ell_p$-sparse
signals whose sorted coefficients obey roughly the same ordering~\cite{modelcs}.
This condition could perhaps be enforced as an $\ell_p$ constraint
on the composite signal
$$
\left\{\sum_{\j=1}^\J |x_\j(1)|,\ \ \sum_{\j=1}^\J |x_\j(2)|,\
\dots,\ \ \sum_{\j=1}^\J |x_\j(\N)|\right\}.
$$

Second, (random)
measurements are real numbers; quantization gradually
degrades the recovery quality as the quantization becomes
coarser \cite{bitscs,quantsparse,goyalquant}.  Moreover, in many practical
situations some amount of measurement noise will corrupt the $\{x_\j\}$,
making them not exactly sparse in any basis. While characterizing
these effects and the resulting rate-distortion consequences in
the DCS setting are topics for future work, there has been work in
the single-signal CS literature that we should be able to
leverage, including variants of Basis Pursuit with
Denoising~\cite{DonohoECS,CandesSSR},
robust iterative recovery algorithms~\cite{HauptNowak}, CS noise sensitivity
analysis \cite{CandesCS,bitscs}, the Dantzig
Selector~\cite{CandesDS}, and one-bit CS~\cite{onebit}.

Third, in some applications, the linear program
associated with some DCS decoders (in JSM-1 and JSM-3) could prove
too computationally intense.  As we saw in JSM-2, efficient
iterative and greedy algorithms could come to the rescue, but these
need to be extended to the multi-signal case. Recent results on recovery
from a union of subspaces give promise for efficient, model-based
algorithms~\cite{modelcs}.

Finally, we focused our theory on models that assign common and
innovation components to the signals in the ensemble. Other models
tailored to specific applications can be posed; for example, in
hyperspectral imaging applications, it is common to obtain strong
correlations only across spectral slices within a certain neighborhood.
It would be then appropriate to pose a common/innovation model with
separate common components that are localized to a subset of the
spectral slices obtained. Results similar to those obtained in
Section~\ref{sec-theory} are simple to derive for models with full-rank
location matrices.

\bigskip

\appendix

\section{Proof of Theorem~\ref{theo:kplusone}}
\label{ap:kplusone}
\qq

Statement 2 is an application of the achievable bound of Theorem~\ref{theo:dwtb_tight_achievable_13} to the case of $J=1$ signal.
It remains then to prove Statements 1 and 3.

{\bf Statement 1 (Achievable, $\M \ge 2\K$):} We first note that,
if $\K \ge \N/2$, then with probability one, the matrix $\Phi$ has
rank $\N$, and there is a unique (correct) recovery. Thus we
assume that $\K < \N/2$.
With probability one, all subsets of up to $2\K$ columns drawn from
$\Phi$ are linearly independent. Assuming this holds, then for two
index sets $\Omega \neq \hatomega$ such that $|\Omega| =
|\hatomega| = \K$, $\mathrm{colspan}( \Phi_\Omega) \cap
\mathrm{colspan}(\Phi_\hatomega)$ has dimension equal to the
number of indices common to both $\Omega$ and $\hatomega$. A
signal projects to this common space only if its coefficients are
nonzero on exactly these (fewer than $\K$) common indices; since
$\|\theta\|_0 = K$, this does not occur. Thus every $\K$-sparse
signal projects to a unique point in $\real^\M$.

{\bf Statement 3 (Converse, $\M \le \K$):} If $\M < \K$,
there is insufficient information in the vector $y$ to recover the
$\K$ nonzero coefficients of $\theta$; thus we assume $\M = \K$.
In this case, there is a single explanation for the measurements
only if there is a single set $\Omega$ of $\K$ linearly
independent columns {\em and} the nonzero indices of $\theta$ are
the elements of $\Omega$. Aside from this pathological case, the
rank of subsets $\Phi_\hatomega$ will generally be less than $\K$
 --- which would prevent robust recovery of signals supported on
$\hatomega$, or will be equal to $\K$ --- which would give ambiguous
solutions among all such sets $\hatomega$. \qed

\section{Proof of Theorem~\ref{theo:achstep1}}
\label{app:achstep1}
We let
\begin{equation}
\numcols := K_C + \sum_{j \in \Lambda} K_j \label{eq:numcols}
\end{equation}
denote the number of columns in $P$.
Because $P \in \mathcal{P}_F(X)$, there exists $\Theta \in
\real^\numcols$ such that $X = P \Theta$. Because $Y = \Phi X$,
$\Theta$ is a solution to $Y = \Phi P \Theta$. We will argue that,
with probability one over $\Phi$,
$$
\Upsilon := \Phi P
$$
has rank
$\numcols$, and thus $\Theta$ is the unique solution to the equation
$Y =\Phi P \Theta =  \Upsilon \Theta$.

We recall that, under our common/innovation model, $P$ has the
form (\ref{eq:pmatrix}),
where $P_C$ is an $N \times K_C$ submatrix of the $N \times N$
identity, and each $P_j$, $j \in \Lambda$, is an $N \times K_j$
submatrix of the $N \times N$ identity.
To prove that $\Upsilon$ has rank $\numcols$, we will require the
following lemma, which we prove in Appendix~\ref{app:matching}.

\begin{LEMM}
If (\ref{eq:achCondition1}) holds, then there exists a mapping $\mathcal{C} :
\{1,2,\dots,K_C\} \rightarrow \Lambda$, assigning each element of the
common component to one of
the sensors, such that for each $\Gamma \subseteq \Lambda$,
\begin{equation}
\sum_{j \in \Gamma} M_j \ge \sum_{j \in \Gamma} K_j + \sum_{k=1}^{K_C}
1_{\mathcal{C}(k) \in \Gamma}
\label{eq:matching}
\end{equation}
and such that for each $k \in \{1,2,\dots,K_C\}$, the
$k^\mathrm{th}$ column of $P_C$ is not a column of
$P_{\mathcal{C}(k)}$.

\label{lemma:matching}
\end{LEMM}

Intuitively, the existence of such a mapping suggests that ($i$)
each sensor has taken enough measurements to cover its own
innovation (requiring $K_j$ measurements) and perhaps some
of the common component, ($ii$) for any $\Gamma \subseteq
\Lambda$, the sensors in $\Gamma$ have collectively taken enough
extra measurements to cover the requisite $K_C(\Gamma,P)$ elements
of the common component, and ($iii$) the extra measurements are
taken at sensors where the common and innovation components do not
overlap.
Formally, we will use the existence of such a mapping to prove that
$\Upsilon$ has rank $\numcols$.

We proceed by noting that $\Upsilon$ has the form
$$
\Upsilon = \left[
\begin{array} {ccccc}
\Phi_1 P_C & \Phi_1 P_1 & \bf{0} & \dots & \bf{0} \\
\Phi_2 P_C & \bf{0} & \Phi_2 P_2 & \dots & \bf{0} \\
\vdots & \vdots & \vdots & \ddots & \vdots \\
\Phi_J P_C & \bf{0} & \bf{0} & \dots & \Phi_J P_J
\end{array}
\right],
$$
where each $\Phi_j P_C$ (respectively, $\Phi_j P_j$) is an $M_j
\times K_C$ (respectively, $M_j \times K_j$) submatrix of $\Phi_j$
obtained by selecting columns from $\Phi_j$ according to the nonzero
entries of $P_C$ (respectively, $P_j$). In total, $\Upsilon$ has
$\numcols$ columns (\ref{eq:numcols}). To argue that $\Upsilon$ has
rank $\numcols$, we will consider a sequence of three matrices
$\Upsilon_0$, $\Upsilon_1$, and $\Upsilon_2$ constructed from small
modifications to $\Upsilon$.

We begin by letting $\Upsilon_0$ denote the ``partially zeroed''
matrix obtained from $\Upsilon$ using the following construction.
We first let $\Upsilon_0 = \Upsilon$ and then make the following
adjustments:
\begin{enumerate}
\item Let $k = 1$.
\item For each $j$ such that $P_j$ has a column that matches column
$k$ of $P_C$ (note that by Lemma~\ref{lemma:matching} this cannot
happen if $\mathcal{C}(k) = j$), let $k'$ represent the column index
of the full matrix $P$ where this column of $P_j$ occurs. Subtract
column $k'$ of $\Upsilon_0$ from column $k$ of $\Upsilon_0$. This
forces to zero all entries of $\Upsilon_0$ formerly corresponding to
column $k$ of the block $\Phi_j P_C$.
\item If $k < K_C$, add one to $k$ and go to step 2.
\end{enumerate}
The matrix $\Upsilon_0$ is identical to $\Upsilon$ everywhere except
on the first $K_C$ columns, where any portion of a column
overlapping with a column of $\Phi_j P_j$ to its right has been set
to zero.
Thus, $\Upsilon_0$ satisfies the next two properties, which
will be inherited by matrices $\Upsilon_1$ and $\Upsilon_2$ that we
subsequently define:
\begin{enumerate}
\item[\textsf{P1.}] Each entry of $\Upsilon_0$ is either zero or a Gaussian random
variable.
\item[\textsf{P2.}] All Gaussian random variables in $\Upsilon_0$ are i.i.d.
\end{enumerate}
Finally, because $\Upsilon_0$ was constructed only by subtracting
columns of $\Upsilon$ from one another,
\begin{equation}
\rank{\Upsilon_0} = \rank{\Upsilon}.
\label{eq:rank}
\end{equation}

We now let $\Upsilon_1$ be the matrix obtained from $\Upsilon_0$
using the following construction.
For each $j\in\Lambda$, we select $K_j + \sum_{k=1}^{K_C}
1_{\mathcal{C}(k) = j}$ arbitrary rows from the portion of
$\Upsilon_0$ corresponding to sensor $j$.
Using~(\ref{eq:numcols}), the resulting matrix $\Upsilon_1$ has
$$
\sum_{j \in \Lambda} \left(K_j + \sum_{k=1}^{K_C}
1_{\mathcal{C}(k) = j} \right) = \sum_{j \in \Lambda}
K_j + K_C  = \numcols
$$
rows. Also, because $\Upsilon_1$ was obtained by selecting a subset
of rows from $\Upsilon_0$, it has $\numcols$ columns
and satisfies
\begin{equation}
\rank{\Upsilon_1} \le \rank{\Upsilon_0}.
\label{eq:rank2}
\end{equation}

We now let $\Upsilon_2$ be the $\numcols \times \numcols$ matrix
obtained by permuting columns of $\Upsilon_1$ using the following
construction:
\begin{enumerate}
\item Let $\Upsilon_2 = [~]$, and let $j = 1$.
\item For each $k$ such that $\mathcal{C}(k) = j$, let
$\Upsilon_1(k)$ denote the $k^\mathrm{th}$ column of $\Upsilon_1$,
and concatenate $\Upsilon_1(k)$ to $\Upsilon_2$, i.e., let
$\Upsilon_2 \leftarrow [\Upsilon_2 ~ \Upsilon_1(k)]$. There are
$\sum_{k=1}^{K_C} 1_{\mathcal{C}(k) = j}$ such columns.
\item Let $\Upsilon_1'$ denote the
columns of $\Upsilon_1$ corresponding to the entries of $\Phi_j P_j$ (the
innovation components of sensor $j$), and concatenate $\Upsilon_1'$
to $\Upsilon_2$, i.e., let $\Upsilon_2 \leftarrow [\Upsilon_2 ~
\Upsilon_1']$. There are $K_j$ such columns.
\item If $j < J$, let $j \leftarrow j+1$ and go to Step 2.
\end{enumerate}
Because $\Upsilon_1$ and $\Upsilon_2$ share the same columns up to
reordering, it follows that
\begin{equation}
\rank{\Upsilon_2} = \rank{\Upsilon_1}.
\label{eq:rank3}
\end{equation}
Based on its dependency on $\Upsilon_0$, and following from
Lemma~\ref{lemma:matching}, the square matrix $\Upsilon_2$ meets
properties \textsf{P1} and \textsf{P2} defined above in addition to
a third property:
\begin{enumerate}
\item[\textsf{P3.}] All diagonal entries of $\Upsilon_2$ are Gaussian random
variables.
\end{enumerate}
This follows because for each $j$,
$K_j +
\sum_{k=1}^{K_C} 1_{\mathcal{C}(k) = j}$ rows of $\Upsilon_1$ are
assigned in its construction, while
$K_j +
\sum_{k=1}^{K_C} 1_{\mathcal{C}(k) = j}$ columns of $\Upsilon_2$ are
assigned in its construction. Thus, each diagonal element of
$\Upsilon_2$ will either be an entry of some $\Phi_j P_j$, which
remains Gaussian throughout our constructions, or it will be an entry
of some $k^\mathrm{th}$ column of some $\Phi_j P_C$ for which
$\mathcal{C}(k) = j$. In the latter case, we know by
Lemma~\ref{lemma:matching} and the construction of $\Upsilon_0$ that
this entry remains Gaussian throughout our constructions.

Having identified these three properties satisfied by $\Upsilon_2$,
we will prove by induction that, with probability one over $\Phi$,
such a matrix has full rank.

\begin{LEMM}
Let $\Upsilon^{(d-1)}$ be a $(d-1) \times (d-1)$ matrix having full
rank. Construct a $d \times d$ matrix  $\Upsilon^{(d)}$ as follows:
$$
\Upsilon^{(d)} := \left[ \begin{array}{cc} \Upsilon^{(d-1)} & v_1 \\
v_2^t & \omega \end{array} \right]
$$
where $v_1, v_2 \in \real^{d-1}$ are vectors with each entry being
either zero or a Gaussian random variable, $\omega$ is a Gaussian
random variable, and all random variables are i.i.d.\ and
independent of $\Upsilon^{(d-1)}$. Then with probability one,
$\Upsilon^{(d)}$ has full rank.
\label{lemma:inductionPhi}
\end{LEMM}

Applying Lemma~\ref{lemma:inductionPhi} inductively $\numcols$
times, the success probability remains one. It follows that with
probability one over $\Phi$, $\rank{\Upsilon_2} = \numcols$. Combining
this last result with (\ref{eq:rank}-\ref{eq:rank3}),
we obtain $\rank{\Upsilon} = \numcols$ with probability one over $\Phi$.
It remains to prove Lemma~\ref{lemma:inductionPhi}.

\noindent {\bf Proof of Lemma~\ref{lemma:inductionPhi}:} When $d =
1$, $\Upsilon^{(d)} = [\omega]$, which has full rank if and only if
$\omega \neq 0$, which occurs with probability one.

When $d > 1$, using expansion by minors, the determinant of
$\Upsilon^{(d)}$ satisfies
$$
\det(\Upsilon^{(d)}) = \omega \cdot \det(\Upsilon^{(d-1)}) + C,
$$
where $C = C(\Upsilon^{(d-1)}, v_1, v_2)$ is independent of
$\omega$. The matrix $\Upsilon^{(d)}$  has full rank if and only if
$\det(\Upsilon^{(d)}) \neq 0$, which is satisfied if and only if
$$
\omega \neq \frac{-C}{\det(\Upsilon^{(d-1)})}.
$$
By assumption, $\det(\Upsilon^{(d-1)}) \neq 0$ and $\omega$ is a
Gaussian random variable that is independent of $C$ and
$\det(\Upsilon^{(d-1)})$. Thus, $\omega \neq
\frac{-C}{\det(\Upsilon^{(d-1)})}$ with probability one. \qed

\section{Proof of Lemma~\ref{lemma:matching}}
\label{app:matching}

To prove this lemma, we apply tools from graph theory.

We seek a matching within the graph $G=(V_V,V_M,E)$ from
Figure~\ref{fig:bgraph}, i.e., a subgraph
$(V_V,V_M,\Etilde)$ with $\Etilde \subseteq \Ebar$ that pairs each element of
$V_V$ with a unique element of $V_M$.
Such a matching will immediately give us the desired mapping
$\mathcal{C}$ as follows: for each $k \in \{1,2,\dots,K_C\}
\subseteq V_V$,
we let $(j,m) \in V_M$ denote the single node matched to
$k$ by an edge in $\Ebar$, and we set $\mathcal{C}(k) = j$.

To prove the existence of such a matching within the graph, we
invoke a version of Hall's marriage theorem for bipartite
graphs~\cite{graphs}.
Hall's theorem states that within a bipartite graph $(V_1,V_2,E)$,
there exists a matching that assigns each element of $V_1$ to a
unique element of $V_2$ if for any collection of elements $\Pi
\subseteq V_1$, the set $E(\Pi)$ of neighbors of $\Pi$ in $V_2$ has
cardinality $|E(\Pi)| \ge |\Pi|$.

In the context of our lemma, Hall's condition requires that for any
set of entries in the value vector, $\Pi \subseteq V_V$, the set
$\Ebar(\Pi)$ of neighbors of $\Pi$ in $V_M$ has size $|\Ebar(\Pi)| \ge
|\Pi|$.
We will prove that if (\ref{eq:achCondition1}) is satisfied,
then Hall's condition is satisfied, and thus a matching must exist.

Let us consider an arbitrary set $\Pi \subseteq V_V$. We let
$\Ebar(\Pi)$ denote the set of neighbors of $\Pi$ in $V_M$ joined by
edges in $\Ebar$, and we let $S_\Pi = \{j \in \Lambda: (j,m) \in \Ebar(\Pi)
\mathrm{~for~some~} m\}$.
Thus, $S_\Pi \subseteq \Lambda$ denotes the set of signal indices
whose measurement nodes have edges that connect to $\Pi$.
It follows that $|\Ebar(\Pi)| = \sum_{j \in S_\Pi} M_j$.
Thus, in order to satisfy Hall's condition for $\Pi$, we require
\begin{equation}
\sum_{j \in S_\Pi} M_j \ge |\Pi|. \label{eq:partialhall222}
\end{equation}
We would now like to show that $\sum_{j \in S_\Pi} K_j +K_C(S_\Pi,P) \ge
|\Pi|$, and thus if (\ref{eq:achCondition1}) is satisfied for all
$\Gamma \subseteq \Lambda$, then (\ref{eq:partialhall222}) is
satisfied in particular for $S_\Pi \subseteq \Lambda$.

In general, the set $\Pi$ may contain vertices for both common
components and innovation components. We write $\Pi = \Pi_I \cup
\Pi_C$ to denote the disjoint union of these two sets.

By construction, $|\Pi_I| = \sum_{j \in S_\Pi} K_j$ because we
count all innovations with neighbors in $S_\Pi$, and because
$S_\Pi$ contains all neighbors for nodes in $\Pi_I$. We will also
argue that $K_C(S_\Pi,P) \ge |\Pi_C|$ as follows.
By definition, for a set $\Gamma \subseteq \Lambda$,
$K_C(\Gamma,P)$ counts the number of columns in $P_C$ that also
appear in $P_j$ for all $j \notin \Gamma$. By construction, for each
$k \in \Pi_C$, node $k$ has no connection to nodes $(j,m)$ for $j
\notin S_\Pi$; thus it must follow that the $k^\mathrm{th}$ column
of $P_C$ is present in $P_j$ for all $j \notin S_\Pi$, due to the
construction of the graph $G$. Consequently, $K_C(S_\Pi,P) \geq
|\Pi_C|$.

Thus, $\sum_{j \in S_\Pi} K_j +K_C(S_\Pi,P) \ge |\Pi_I| + |\Pi_C| = |\Pi|$,
and so (\ref{eq:achCondition1}) implies (\ref{eq:partialhall222})
for any $\Pi$, and so Hall's condition is satisfied, and a matching
exists. Because in such a matching a set of vertices in $V_M$ matches to
a set in $V_V$ of lower or equal cardinality, we have in particular that
(\ref{eq:matching}) holds for each $\Gamma \subseteq \Lambda$. \qed

\section{Proof of Theorem~\ref{theo:dwtb_tight_achievable_13}}
\label{app:achstep2}
Given the measurements $Y$ and measurement matrix $\Phi$, we will
show that it is possible to recover some $P\in
\mathcal{P}_F(X)$ and a corresponding vector $\Theta$ such that
$X = P \Theta$ using the following algorithm:
\begin{itemize}
\item Take the last measurement of each sensor for verification, and sum
these $J$ measurements to obtain a single {\em global} test measurement
$\widebar{y}$. Similarly, add the corresponding rows of
$\Phi$ into a single row $\widebar{\phi}$.
\item Group all the
remaining $\sum_{j\in \Lambda} M_j - J $ measurements into a vector
$\widebar{Y}$ and a matrix $\widebar{\Phi}$.
\item For each matrix $P \in \mathcal{P}$:
\begin{itemize}
\item choose a single solution
$\Theta_P$ to $\widebar{Y} = \widebar{\Phi} P \Theta_P$ independently of $\widebar{\phi}$
--- if no solution exists, skip the next two steps;
\item define $X_P = P\Theta_P$;
\item cross-validate: check if $\widebar{y} = \widebar{\phi}X_P$; if so, return the estimate
$(P,\Theta_P)$;
if not, continue with the next matrix.
\end{itemize}
\end{itemize}
We begin by showing that, with probability one over $\Phi$, the algorithm
only terminates when it gets a correct solution --- in other words, that for
each $P \in \mathcal{P}$ the cross-validation measurement $\widebar{y}$ can
determine whether $X_P = X$.
We note that all entries of the vector $\widebar{\phi}$ are i.i.d.\ Gaussian, and
independent from $\widebar{\Phi}$.
Assume for the sake of contradiction that there exists a matrix $P \in \mathcal{P}$
such that  $\widebar{y} = \widebar{\phi}X_{P}$, but $X_P = P
\Theta_{P} \neq X$; this
implies $\widebar{\phi}(X-X_{P}) = 0$, which occurs with probability
zero over
$\Phi$. Thus, if $X_P \ne X$, then $\widebar{\phi} X_P \ne \widebar{y}$ with probability
one over $\Phi$. Since we only need to search over a finite number of matrices
$P \in \mathcal{P}$, cross validation will determine whether each matrix
$P \in \mathcal{P}$ gives the correct solution with probability one.

We now show that there is a matrix in $\mathcal{P}$ for which the algorithm
will terminate with the correct solution.
We know that the matrix $P^* \in \mathcal{P}_F(X) \subseteq \mathcal{P}$
will be part of our search, and that the unique solution $\Theta_{P^*}$ to
$\widebar{Y} = \widebar{\Phi} P^* \Theta_{P^*}$ yields $X = P^* \Theta_{P^*}$ when (\ref{eq:dwtb_loose_achievable_condition_13}) holds for $P^*$, as shown in
Theorem~\ref{theo:achstep1}. Thus, the algorithm will find at least one matrix
$P$ and vector $\Theta_P$ such that $X = P \Theta_P$; when such matrix is
found the cross-validation step will return this solution and end the algorithm.
\qed

\begin{REMA}
Consider the algorithm used in the proof: if the matrices in
$\mathcal{P}$ are sorted by number of columns, then the algorithm is akin to
$\ell_0$-norm minimization on $\Theta$ with an additional cross-validation step.
The $\ell_0$-norm minimization algorithm is known to be optimal for recovery 
of strictly sparse signals from noiseless measurements.
\end{REMA}

\section{Proof of Theorem~\ref{theo:cnv}}
\label{app:cnv}

We let $\numcols$ denote the number of columns in $P$. Because $P
\in \mathcal{P}_F(X)$, there exists $\Theta \in \real^\numcols$ such
that $X = P \Theta$.
Because $Y = \Phi X$, then $\Theta$ is a solution to $Y = \Phi P
\Theta$.
We will argue for $\Upsilon := \Phi P$ that $\rank{\Upsilon} <
\numcols$, and thus there exists $\widehat{\Theta} \neq \Theta$ such
that $Y = \Upsilon \Theta = \Upsilon \widehat{\Theta}$.
Moreover, since $P$ has full rank, it follows that $\widehat{X} := P
\widehat{\Theta} \neq P \Theta = X$.

We let $\Upsilon_0$ be the ``partially zeroed'' matrix obtained from
$\Upsilon$ using the identical procedure detailed in
Appendix~\ref{app:achstep1}.
Again, because $\Upsilon_0$ was constructed only by subtracting
columns of $\Upsilon$ from one another, it follows that
$\rank{\Upsilon_0} = \rank{\Upsilon}$.

Suppose $\Gamma \subseteq \Lambda$ is a set for which
(\ref{eq:cnvCondition1}) holds.
We let $\Upsilon_1$ be the submatrix of $\Upsilon_0$ obtained by
selecting the following columns:
\begin{itemize}
\item For any $k \in \{1,2,\dots,K_C\}$ such that column $k$ of
$P_C$ also appears as a column in all $P_j$ for $j \notin \Gamma$,
we include column $k$ of $\Upsilon_0$ as a column in $\Upsilon_1$.
There are $K_C(\Gamma,P)$ such columns $k$.
\item For any $k \in \{K_C+1,K_C+2,\dots,\numcols\}$ such that
column $k$ of $P$ corresponds to an innovation for some sensor $j
\in \Gamma$, we include column $k$ of $\Upsilon_0$ as a column in
$\Upsilon_1$. There are $\sum_{j \in \Gamma} K_j$ such columns $k$.
\end{itemize}
This submatrix has
$\sum_{j \in \Gamma} K_j+K_C(\Gamma,P)$
columns. Because $\Upsilon_0$ has the same size as  $\Upsilon$, and
in particular has only $\numcols$ columns, then in order to have
that $\rank{\Upsilon_0} = \numcols$, it is necessary that all
$\sum_{j \in \Gamma} K_j+K_C(\Gamma,P)$ columns of $\Upsilon_1$ be
linearly independent.

Based on the method described for constructing
$\Upsilon_0$, it follows that $\Upsilon_1$ is zero for all
measurement rows not corresponding to the set $\Gamma$. Therefore,
consider the submatrix $\Upsilon_2$ of $\Upsilon_1$ obtained
by selecting only the measurement rows corresponding to the set
$\Gamma$. Because of the zeros in $\Upsilon_1$, it follows that
$\rank{\Upsilon_1} = \rank{\Upsilon_2}$. However, since $\Upsilon_2$
has only $\sum_{j \in \Gamma} M_j$ rows, we invoke
(\ref{eq:cnvCondition1}) and have that $\rank{\Upsilon_1} =
\rank{\Upsilon_2} \le \sum_{j \in \Gamma} M_j <
\sum_{j \in \Gamma} K_j+K_C(\Gamma,P)$. Thus, all
$\sum_{j \in \Gamma} K_j+K_C(\Gamma,P)$ columns of $\Upsilon_1$ cannot be
linearly independent, and so $\Upsilon$ does not have full rank.
\qed
\input{jsm1proofs}

\qq
\section{Proof of Theorem~\ref{theo:onepass}}
\label{ap:onepass}\qq

We assume that $\Psi$ is an orthonormal matrix. Like $\Phi_\j$
itself, the matrix $\Phi_\j \Psi$ also has i.i.d.\ $\cl{N}(0,1)$
entries, since $\Psi$ is orthonormal. For convenience, we assume
$\Psi = I_N$. The results presented can be easily extended to a more
general orthonormal matrix $\Psi$ by replacing $\Phi_\j $ with
$\Phi_\j \Psi$.

Assume without loss of generality that $\Omega = \{1,2,\dots,\K\}$
for convenience of notation. Thus, the correct estimates are $\n \le
\K$, and the incorrect estimates are $\n \ge \K+1$. Now consider the
statistic $\xi_\n$ in (\ref{eq:oneshot}). This is the sample mean of
$\J$ i.i.d.\ variables. The variables $\langle y_j,
\phi_{j,n}\rangle^2$ are i.i.d.\ since each $y_j=\Phi_j x_j$, and
$\Phi_j$ and $x_j$ are i.i.d. Furthermore, these variables have a
finite variance.\footnote{In \cite{CLTTR2005}, we evaluate the
variance of $\langle y_\j , \phi_{\j,\n} \rangle^2 $ as
$$
\mbox{Var}[\langle y_\j , \phi_{\j,\n} \rangle^2] = \left\{
\begin{array}{ll}
\M\sigma^4(34\M\K+6\K^2+28\M^2+92\M+48\K+90+2\M^3+2\M\K^2+4\M^2\K), &
\n \in \Omega \\
2\M\K\sigma^4(\M\K+3\K+3\M+6), & \n \notin \Omega.
\end{array}
\right.
$$
For finite $\M$, $\K$ and $\sigma$, the above variance is finite.
}
Therefore, we invoke the Law of Large Numbers (LLN)
to argue that $\xi_n$, which is a sample mean
of $\langle y_j, \phi_{j,n}\rangle^2$,
converges to $E[\langle y_j, \phi_{j,n}\rangle^2]$ as $J$ grows large.
We now compute $E[\langle y_\j ,
\phi_{\j,\n} \rangle^2]$ under two cases. In the first case, we
consider $\n \ge \K+1$ (we call this the ``bad statistics case''),
and in the second case, we consider $\n \le \K$ (``good statistics
case'').

{\bf Bad statistics:}  Consider one of the bad statistics by
choosing $\n=\K+1$ without loss of generality. We have
\begin{eqnarray}
E[\langle y_\j , \phi_{\j,\K+1} \rangle^2] &=& E \left[
\sum_{\n=1}^{\K}x_\j(\n)\langle \phi_{\j,\n}, \phi_{\j,\K+1}\rangle
\right]^2 \nonumber  \\
&=& E\left[\sum_{\n=1}^{\K} x_\j(\n)^2  \langle \phi_
{\j,\n}, \phi_{\j,\K+1}\rangle^2    \right] \nonumber \\
& & + ~ E\left[\sum_{\n=1}^{\K}\sum_{\ell=1, \ell \ne
\n}^{\K}x_\j(\ell)x_\j(\n) \langle \phi_{\j,\ell},
\phi_{\j,\K+1}\rangle \langle \phi_{\j,\n}, \phi_{\j,\K+1}\rangle
\right] \nonumber \\
&=&
\sum_{\n=1}^{\K}E\left[x_\j(\n)^2\right]
E\left[\langle \phi_{\j,\n}, \phi_{\j,\K+1}\rangle^2\right]
\nonumber \\
& & + ~ \sum_{\n=1}^{\K}\sum_{\ell=1, \ell \ne
\n}^{\K}E[x_\j(\ell)]E[x_\j(\n)] E\left[\langle \phi_{\j,\ell},
\phi_{\j,\K+1}\rangle
\langle \phi_{\j,\n}, \phi_{\j,\K+1}\rangle\right], \nonumber
\end{eqnarray}
since the terms are independent. We also have
$E[x_\j(\n)]=E[x_\j(\ell)]=0$, and so
\begin{eqnarray}
E[\langle y_\j , \phi_{\j,\K+1} \rangle^2]
&=&
\sum_{\n=1}^{\K}E\left[x_\j(\n)^2\right]
E\left[\langle
\phi_{\j,\n}, \phi_{\j,\K+1}\rangle^2\right]
\nonumber \\
&=& \sum_{\n=1}^{\K} \sigma^2 E\left[ \langle \phi_{\j,\n},
\phi_{\j,\K+1}\rangle^2\right]. \label{equ:lln:badstats}
\end{eqnarray}
To compute $E\left[\langle \phi_{\j,\n},
\phi_{\j,\K+1}\rangle^2\right]$, let $\phi_{\j,\n}$ be the column
vector $[a_1, a_2, \ldots, a_\M]\trans$, where each element in the
vector is i.i.d.\ $\cl{N}(0,1)$. Likewise, let $\phi_{\j,\K+1}$ be
the column vector $[b_1, b_2, \ldots, b_\M]\trans$ where the elements
are i.i.d.\ $\cl{N}(0,1)$. We have
\begin{eqnarray*}
\langle \phi_{\j,\n} , \phi_{\j,\K+1} \rangle^2 &=&
(a_1b_1 + a_2b_2 + \ldots + a_\M b_\M)^2 \\
&=& \sum_{\m=1}^{\M}a_\m^2b_\m^2  +
2\sum_{\m=1}^{\M-1}\sum_{r=\m+1}^{\M}a_\m a_rb_\m b_r.
\end{eqnarray*}
Taking the expected value, we have
\begin{eqnarray*}
E\left[ \langle \phi_{\j,\n} , \phi_{\j,\K+1} \rangle^2 \right]&=&
E\left[\sum_{\m=1}^{\M}a_\m^2b_\m^2 \right] +  2E\left[\sum_{\m=1}^
{\M-1}\sum_{r=\m+1}^{\M}a_\m a_rb_\m b_r\right] \\
&=& \sum_{\m=1}^{\M}E\left[a_\m^2b_\m^2\right] +
2\sum_{\m=1}^{\M-1}\sum_{r=q+1}^{\M}E\left[a_\m a_rb_\m b_r\right]\\
&=&
\sum_{\m=1}^{\M}E\left[a_\m^2\right] E\left[b_\m^2\right] +
2\sum_{\m=1}^{\M-1}\sum_{r=\m+1}^{\M} E\left[a_\m\right] E\left[a_r
\right]
E\left[b_\m\right] E\left[b_r\right] \\
& & \text{\hspace{43mm}(since the random variables are independent)} \\
&=&
\sum_{\m=1}^{\M} (1) + 0
\text{\hspace{10mm}(since
$E\left[a_\m^2\right]=E\left[b_\m^2\right]=1$
and $E\left[a_\m\right]=E\left[b_\m\right]=0 $)} \\
&=& \M,
\end{eqnarray*}
and thus
\begin{eqnarray}
E\left[ \langle \phi_{\j,\n} , \phi_{\j,\K+1} \rangle^2 \right]&=&\M.
\label{equ:lln:result1}
\end{eqnarray}
Combining this result with (\ref{equ:lln:badstats}), we find that
\begin{eqnarray*}
E[\langle y_\j , \phi_{\j,\K+1} \rangle^2]
= \sum_{\n=1}^{\K} \sigma^2 \M
= \M\K \sigma ^2.
\end{eqnarray*}
Thus we have computed $E[\langle y_\j , \phi_{\j,\K+1} \rangle^2]$ and
can conclude that as $\J$ grows large, the statistic $\xi_{\K+1}$
converges to
\begin{equation}
E[\langle y_\j , \phi_{\j,\K+1} \rangle^2] = \M\K\sigma^2.
\label{equ:lln:badstats1}
\end{equation}

{\bf Good statistics:} Consider one of the good statistics, and without loss
of generality choose $\n=1$. Then, we have
\begin{eqnarray}
E[\langle y_\j , \phi_{\j,1} \rangle^2] &=&
E\left[\left(
x_\j(1)\|\phi_{\j,1}\|^2 + \sum_{\n=2}^{\K}x_\j(\n)\langle\phi_{\j,\n},
\phi_{\j,1} \rangle  \right)^2\right]
\nonumber \\
&=& E\left[\left(x_\j(1)\right)^2\|\phi_{\j,1}\|^4  \right] +
E\left[\sum_{\n=2}^{\K}x_\j(\n)^2\langle\phi_{\j,\n},
\phi_{\j,1} \rangle ^2\right]
\nonumber \\
& & \text{\hspace{55mm}(all other cross terms have zero expectation)}
\nonumber \\
&=& E\left[x_\j(1)^2\right] E\left[\|\phi_{\j,1}\|^4\right]   +
\sum_{\n=2}^{\K}E\left[x_\j(\n)^2\right]
E\left[\langle\phi_{\j,\n}, \phi_{\j,1}
\rangle^2\right] \text{\hspace{6mm} (by independence)}
\nonumber \\
&=& \sigma^2 E\left[\|\phi_{\j,1}\|^4\right]   +
\sum_{\n=2}^{\K}\sigma^2 E\left[\langle\phi_{\j,\n}, \phi_{\j,1}
\rangle^2\right]. \label{equ:lln:goodstats2}
\end{eqnarray}
Extending the result from (\ref{equ:lln:result1}), we can show that
$E\langle\phi_{\j,\n}, \phi_{\j,1} \rangle ^2 = \M$. Using this
result in (\ref{equ:lln:goodstats2}),
\begin{eqnarray}
E[\langle y_\j , \phi_{\j,1} \rangle^2]  &=& \sigma^2E\|\phi_{\j,1}\|^4
+ \sum_{\n=2}^{\K}\sigma^2\M. \label{equ:lln:goodstats1}
\end{eqnarray}
To evaluate $E\left[\|\phi_{\j,1}\|^4\right]$, let $\phi_{\j,1}$ be
the column
vector $[c_1, c_2, \ldots, c_\M]\trans$, where the elements of the
vector are random $\mathcal{N}(0,1)$.
Define the random variable $Z=\|  \phi_{\j,1}
\|^2=\sum_{\m=1}^{\M}c_\m^2$. Note that ($i$)
$E\left[\|\phi_{\j,1}\|^4\right]=E\left[Z^2\right]$ and ($ii$)
$Z$ is chi-squared distributed with $\M$
degrees of freedom. Thus,
$E\left[\|\phi_{\j,1}\|^4\right]=E\left[Z^2 \right]=\M(\M+2)$.
Using this result in (\ref{equ:lln:goodstats1}), we have
\begin{eqnarray*}
E[\langle y_\j , \phi_{\j,1} \rangle^2]
&=& \sigma^2\M(\M+2) + (\K-1)\sigma^2\M \\
&=& \M(\M+\K+1)\sigma^2.
\end{eqnarray*}
We have computed the variance of $\langle y_\j , \phi_{\j,1} \rangle$
and can conclude that as $\J$ grows large, the statistic $\xi_1$
converges to
\begin{equation}
E[\langle y_\j , \phi_{\j,1} \rangle^2] = (\M+\K+1)\M\sigma^2.
\label{equ:lln:goodstats3}
\end{equation}

{\bf Conclusion:} From (\ref{equ:lln:badstats1}) and
(\ref{equ:lln:goodstats3}) we conclude that
$$
\lim_{\J \rightarrow \infty} \xi_\n = E[\langle y_\j , \phi_{\j,\n}
\rangle^2] = \left\{
\begin{array}{ll}
(\M+\K+1)\M\sigma^2, & \n \in \Omega \\
\K\M\sigma^2, & \n \notin \Omega.
\end{array}
\right.
$$
For any $\M \ge 1$, these values are distinct --- their ratio is  $\frac {\M + \K + 1}{\K}$.
Therefore, as $\J$ increases we can distinguish between the two
expected values of $\xi_\n$ with overwhelming probability.
\qed

\qq
\section{Proof of Theorem~\ref{theo:nonsparse}}
\label{ap:jsm3} \qq
%
Our proof has two parts.
First we argue that $\lim_{\J \rightarrow
\infty} \widehat{z}_C = z_C$. Then we show that this implies
vanishing probability of error in recovering each innovation $z_\j$.

\noindent{\bf Part 1:} We can write our estimate as
\begin{equation*}
\widehat{z}_C = \frac{1}{\J} \widehat{\Phi}\trans y  =
\frac{1}{\J}\sum_{\j=1}^\J \widehat{\Phi}_j\trans y_j =
\frac{1}{\J}\sum_{\j=1}^\J\frac{1}{\M_\j \sigma_\j^2}\Phi_\j\trans
\Phi_\j x_\j = \frac{1}{\J} \sum_{\j=1}^\J
\frac{1}{\M_\j\sigma_\j^2} \sum_{\m=1}^{\M_\j}
(\phi_{\j,\m}^R)\trans \phi_{\j,\m}^R x_j,
\end{equation*}
where
$\phi_{\j,\m}^R$ denotes the $\m$-th row of $\Phi_\j$, that is,
the $\m$-th measurement vector for node $\j$. Since the elements of
each $\Phi_\j$ are Gaussians with variance $\sigma_\j^2$, the
product $(\phi_{\j,\m}^R)\trans \phi_{\j,\m}^R$ has the property
\[
E[(\phi_{\j,\m}^R)\trans \phi_{\j,\m}^R] = \sigma_\j^2I_\N.
\]
It follows that
\[
E[(\phi_{\j,\m}^R)\trans \phi_{\j,\m}^R x_\j] = \sigma_\j^2E[x_\j] =
\sigma_\j^2E[z_C + z_\j] =
\sigma_\j^2z_C
\]
and, similarly, that
\[
E\left[\frac{1}{\M_\j\sigma_\j^2}
\sum_{\m=1}^{\M_\j}(\phi_{\j,\m}^R)\trans \phi_{\j,\m}^R
x_\j\right] = z_C.
\]
Thus, $\widehat{z}_C$ is a sample mean of $\J$ independent random
variables with mean $z_C$. From the law of large numbers, we conclude
that $\lim_{\J \rightarrow \infty} \widehat{z}_C = z_C$.

\noindent{\bf Part 2:} Consider recovery of the innovation $z_\j$
from the adjusted measurement vector $\widehat{y_\j} = y_\j -
\Phi_\j \widehat{z}_C$. As a recovery scheme, we consider a
combinatorial search over all $\K$-sparse index sets drawn from
$\{1,2,\dots,\N\}$. For each such index set $\Omega'$, we compute
the distance from $\widehat{y}$ to the column span of
$\Phi_{\j,\Omega'}$, denoted by $d(\widehat{y},
\mathrm{colspan}(\Phi_{j,\Omega'}))$, where $\Phi_{\j,\Omega'}$ is
the matrix obtained by sampling the columns $\Omega'$ from
$\Phi_\j$. (This distance can be measured using the pseudoinverse of
$\Phi_{\j,\Omega'}$.)

For the correct index set $\Omega$, we know that
$d(\widehat{y_\j},\mathrm{colspan}(\Phi_{\j,\Omega})) \rightarrow 0$ as
$\J \rightarrow \infty$. For any other index set $\Omega'$, we know
from the proof of Theorem~\ref{theo:kplusone} that
$d(\widehat{y_\j},\mathrm{colspan}(\Phi_{\j,\Omega'})) > 0$. Let
\vspace{-1mm}
\[
\zeta := \min_{\Omega' \neq \Omega}
d(\widehat{y_\j},\mathrm{colspan}( \Phi_{i,\Omega'})).
\vspace{-1mm}
\]
With probability one, $\zeta > 0$. Thus for sufficiently large
$\J$, we will have
$d(\widehat{y_\j},\mathrm{colspan}(\Phi_{\j,\Omega})) < \zeta/2$,
and so the correct index set $\Omega$ can be correctly
identified. Since $\lim_{\J
\rightarrow \infty} \widehat{z}_C = z_C$, the innovation estimates
$\widehat{z_j} = z_j$ for each $j$ and for $J$ large enough. \qed

\vspace{-2mm}
\section*{Acknowledgments}
\vspace{-1mm}

Thanks to Emmanuel
Cand\`{e}s, Albert Cohen, Ron DeVore, Anna Gilbert,
Illya Hicks, Robert Nowak, Jared Tanner, and Joel Tropp for informative and
inspiring conversations. Special thanks to
to Mark Davenport for a thorough critique of the manuscript.
MBW thanks his former affiliated institutions Caltech and the University of
Michigan, where portions of this work were performed.
Final thanks to Ryan King for supercharging our computational capabilities.

\vspace{-2mm}
\footnotesize
\bibliography{DCS112005}
\bibliographystyle{ieeetr}

\end{document}

%% file: jsm1section.tex
\qq
\subsection{Recovery strategies for sparse common + innovations (JSM-1)} \qq
\label{sec-mrr}

We first characterize the sparse common signal and innovations model JSM-1 from Section \ref{sec-ds1}.
For simplicity, we limit our description to
$J=2$ signals, but describe extensions to multiple signals as needed.

\qq
\subsubsection{Measurement bounds for joint recovery}
\label{subsec:jsm1:ell_0} \qq

Under the JSM-1 model,
separate recovery of the signal $x_j$ via $\ell_0$-norm minimization would
require $\K_{\mathrm{joint}}(\{j\}) +1 = \K_{\mathrm{cond}}(\{j\}) +1 = K_C+\K_j-K_C(\Lambda\setminus\{j\})+1$ measurements, where $K_C(\Lambda\setminus\{j\})$ accounts for 
sparsity reduction due to overlap between $z_C$ and $z_j$. We apply
Theorem~\ref{theo:dwtb_tight_achievable_13} to the JSM-1 model to obtain
the corollary below. To address the possibility of sparsity reduction, we denote
by $K_R$ the number of indices in which the common component $z_C$ and
all innovation components $z_j$, $j \in \Lambda$ overlap; this results in sparsity
reduction for the common component.

\begin{COROLLARY}
\label{cor:dcs1ell0achievable} Assume the measurement matrices
$\{\Phi_\j\}_{j \in \Lambda}$ contain i.i.d.\ Gaussian entries.
Then the signal ensemble $X$ can be recovered 
with probability one if the following conditions hold:
\begin{subequations}
\begin{eqnarray}
\sum_{\j \in \Gamma} \M_\j &\ge& \left(\sum_{\j \in \Gamma} \K_\j\right) +
\K_C(\Gamma) + |\Gamma|, ~~~ \Gamma \ne \Lambda,
\label{eq:D_WTB:achievable1} \nonumber \\
\sum_{\j\in\Lambda} \M_\j &\ge& \K_C + \left(\sum_{\j \in \Lambda} \K_\j\right)
+ J - K_R. \nonumber
\label{eq:D_WTB:achievable}
\end{eqnarray}
\end{subequations}
\end{COROLLARY}
Our joint recovery scheme provides a significant savings in measurements,
because the common component can be measured as part of any of the $J$ signals.

\qq
\subsubsection{Stochastic signal model for JSM-1}
\label{subsec:sigmodel} \qq

To give ourselves a firm footing for analysis, in the remainder of
Section~\ref{sec-mrr} we use a stochastic process for JSM-1 signal generation.
This framework provides an information theoretic setting where we can scale the
size of the problem and investigate which measurement rates enable recovery.
We generate the common and innovation components as follows. For
$n\in \{1,\ldots,N\}$ the decision whether $z_C(n)$ and $z_j(n)$ is zero or not is
an i.i.d.\ Bernoulli process, where the probability of a nonzero value is
given by parameters denoted $S_C$ and $S_j$, respectively.
The values of the nonzero coefficients are then generated from an i.i.d.\ Gaussian
distribution. The outcome of this process is that $z_C$ and $z_j$ have
sparsities $K_C \sim \mbox{Binomial}(N,S_C)$ and $K_j \sim \mbox{Binomial}(N,S_j)$.
The parameters $S_j$ and $S_C$ are {\em sparsity rates} controlling the
random generation of each signal. Our model resembles the Gaussian spike 
process~\cite{WeidmannV99}, which is a limiting case of a Gaussian mixture 
model.

{\bf Likelihood of sparsity reduction and overlap:} This stochastic model can yield signal
ensembles for which the corresponding generating matrices $P$ allow for sparsity
reduction; specifically, there might be overlap between the supports of the common
component $z_C$ and all the innovation components $z_j$, $j \in \Lambda$. For $J=2$,
the probability that a given index is present in all supports is
$S_R := S_C S_1 S_2$.
Therefore, the distribution of the cardinality of this overlap is $K_R \sim \mbox{Binomial}
(N,S_R)$.
We must account for the reduction obtained from the removal of the corresponding number
of columns from the location matrix $P$ when the total number of measurements $M_1+M_2$ is considered. In the same way we can show that the distributions for the
number of indices in the overlaps required by Corollary~\ref{cor:dcs1ell0achievable} are
$K_C(\{1\}) \sim \mbox{Binomial} (N,S_{C,\{1\}})$ and $K_C(\{2\}) \sim \mbox{Binomial}
(N,S_{C,\{2\}})$, where 
$
S_{C,\{1\}} := S_C (1-S_1) S_2 ~~~\mbox{and}~~~S_{C,\{2\}} := S_C  S_1 (1-S_2).
$

{\bf Measurement rate region}: To characterize DCS recovery performance, 
we introduce a {\em
measurement rate region}. We define the measurement
rate $\R_j$ in an asymptotic manner as
\begin{equation}
\R_j := \lim_{\N\rightarrow\infty} \frac{\M_j}{N},~j \in \Lambda \nonumber.
\end{equation}
Additionally, we note that
\begin{equation}
\lim_{\N\rightarrow\infty} \frac{\K_C}{N} = S_C~\textrm{and}~\lim_{\N\rightarrow\infty} \frac{\K_j}{N}  = S_j,~j \in \Lambda \nonumber
\end{equation}
Thus, we also set $S_{X_j} = S_C + S_j - S_CS_j$, $j \in \{1,2\}$.
For a measurement rate pair $(\R_1,\R_2)$ and sources $X_1$ and
$X_2$, we evaluate whether we can recover the signals with
vanishing probability of error as $\N$ increases. In this case, we say that
the measurement rate pair is {\em achievable}.

For jointly sparse signals under JSM-1, separate
recovery via $\ell_0$-norm minimization would require a measurement rate
$R_j=S_{X_j}$. Separate recovery via $\ell_1$-norm minimization would
require an overmeasuring factor $c(S_{X_j})$, and thus the measurement
rate would become $R_j = S_{X_j} \cdot c(S_{X_j})$. To improve upon these
figures, we adapt the standard machinery of CS to the joint recovery
problem.

\qq
\subsubsection{Joint recovery via $\ell_1$-norm minimization}
\label{subsec:jsm1_ell1} \qq

As discussed in Section~\ref{subsec-CSrecovery}, solving an $\ell_0$-norm
minimization is NP-hard, and so in practice we must relax our $\ell_0$
criterion in order to make the solution tractable. 
We now study what penalty must be paid for $\ell_1$-norm recovery of jointly
sparse signals.
Using the vector and frame
\begin{equation}
Z := \left[ \begin{array}{c} z_C \\ z_1 \\ z_2
\end{array}\right]~~~\mathrm{and}~~~\widetilde{\Phi} := \left[ \begin{array}{ccc} \Phi_1 & \Phi_1 & 0 \\
\Phi_2 & 0 & \Phi_2
\end{array} \right],
\end{equation}
we can represent the concatenated measurement vector $Y$ sparsely using the
concatenated coefficient vector $Z$, which contains $\K_C+\K_1+\K_2-\K_R$
nonzero coefficients, to obtain $Y = \widetilde{\Phi}Z$. 
With sufficient overmeasuring, we have seen experimentally that it is possible to recover a
vector $\widehat{Z}$, which yields $x_j = \widehat{z}_C + \widehat{z}_j$, $j=1,2$,
by solving the weighted $\ell_1$-norm minimization
\begin{equation}
\widehat{Z} = \arg\min \gamma_C ||z_C||_1 + \gamma_1||z_1||_1 + \gamma_2||z_2||_1~~\mbox{~s.t.~}y=\widetilde{\Phi}Z,
\label{eq:gamma}
\end{equation}
where $\gamma_C,\gamma_1,\gamma_2 \ge 0$. We call this the {\em
$\gamma$-weighted $\ell_1$-norm formulation}; our numerical results
(Section~\ref{subsec-examplesjsm1} and our technical report~\cite{DCSTR06})
indicate a reduction in the requisite number of measurements via this enhancement.
If $\K_1=\K_2$ and $M_1=M_2$, then without loss of generality we set
$\gamma_1=\gamma_2=1$ and numerically search for the best parameter $\gamma_C$.
We discuss the asymmetric case with $K_1=K_2$
and $M_1 \neq M_2$ in the technical report~\cite{DCSTR06}.

\qq
\subsubsection{Converse bound on performance of $\gamma$-weighted
$\ell_1$-norm minimization} \label{subsec-ell_1_bounds} \qq

We now provide a converse bound that describes what
measurement rate pairs {\em cannot} be achieved via the $\gamma$-weighted $\ell_1$-norm minimization.
Our notion of a converse focuses on the
setting where each signal $x_\j$ is measured via multiplication by the
$\M_\j$ by $N$ matrix $\Phi_j$ and joint recovery is performed via
our $\gamma$-weighted $\ell_1$-norm formulation (\ref{eq:gamma}). Within
this setting, a converse region is a set of measurement rates for
which the recovery fails with overwhelming probability as $N$ increases.

We assume that $\J=2$ sources have innovation sparsity rates that satisfy
$S_1=S_2=S_I$. Our first result, proved in Appendix~\ref{ap:lem_con},
provides deterministic necessary conditions to recover the components
$z_C$, $z_1$, and $z_2$, using the $\gamma$-weighted
$\ell_1$-norm formulation (\ref{eq:gamma}). We note that the lemma holds for
all such combinations of components that generate the same signals
$x_1 = z_C + z_1$ and $x_2 = z_C + z_2$.

\begin{LEMM} \label{lem:con}
Consider any $\gamma_C$, $\gamma_1$, and $\gamma_2$ in the
$\gamma$-weighted $\ell_1$-norm formulation (\ref{eq:gamma}). The
components $z_C$, $z_1$, and $z_2$
can be recovered using measurement matrices $\Phi_1$ and $\Phi_2$
only if (i) $z_1$ can be recovered via $\ell_1$-norm minimization (\ref{eq:L1})
using $\Phi_1$ and measurements
$\Phi_1z_1$; (ii) $z_2$ can be recovered via
$\ell_1$-norm minimization using $\Phi_2$ and measurements
$\Phi_2z_2$; and (iii) $z_C$ can be recovered
via $\ell_1$-norm minimization using the joint matrix $[\Phi_1\trans
~~ \Phi_2\trans]\trans$ and measurements
$[\Phi_1\trans~~\Phi_2\trans]\trans z_C$.
\end{LEMM}

Lemma~\ref{lem:con} can be interpreted as follows. If $M_1$ and
$M_2$ are not large enough individually, then the innovation
components $z_1$ and $z_2$ cannot be
recovered. This implies a converse bound on the individual
measurement rates $R_1$ and $R_2$. Similarly, combining
Lemma~\ref{lem:con} with the converse bound of
Theorem~\ref{theo:bpconstant} for single-source $\ell_1$-norm minimization
of the common component $z_C$ implies a lower bound on the
sum measurement rate $R_1+R_2$.

{\bf Anticipated converse:}
As shown in Corollary~\ref{cor:dcs1ell0achievable}, for indices $n$ such that
$x_1(n)$ and $x_2(n)$ differ and are nonzero, each sensor must take
measurements to account for one of the two coefficients.
In the case where $S_1=S_2=S_I$, the joint sparsity rate is $S_C+2S_I-S_CS_I^2$. We define the measurement function 
$c'(S) := S\cdot c(S)$ based on Donoho and Tanner's oversampling 
factor $c(S)$ (Theorem~\ref{theo:bpconstant}).
It can be shown that the function $c'(\cdot)$ is concave;
in order to minimize the sum rate bound, we ``explain" as many of the sparse
coefficients in one of the signals and as few as possible in the other. From
Corollary~\ref{cor:dcs1ell0achievable}, we have
$R_1, R_2 \ge S_I+S_CS_I-S_CS_I^2$.
Consequently, one of the signals must ``explain" this sparsity rate,
whereas the other signal must explain the rest:
\begin{eqnarray*}
[S_C+2S_I-S_CS_I^2] - [S_I+S_CS_I-S_CS_I^2] = S_C+S_I-S_CS_I.
\end{eqnarray*}
Unfortunately, the derivation of $c'(S)$ relies on Gaussianity of the
measurement matrix, whereas in our case $\Phi$ has a block matrix form. 
Therefore, the following conjecture
remains to be proved rigorously.

\begin{CONJECTURE}
\label{conj:ell_1_conv} Let $\J=2$ and fix the sparsity rate of the
common component $S_C$ and the innovation sparsity rates
$S_1=S_2=S_I$. Then the following conditions on the
measurement rates are necessary to enable recovery with
probability one:
\begin{eqnarray*}
\R_j &\geq& c'\left( S_I+S_CS_I-S_CS_I^2 \right) ,~j=1,2, \\
\R_1+\R_2 &\geq& c'\left( S_I+S_CS_I-S_CS_I^2 \right) + c'\left(
S_C+S_I-S_CS_I \right).
\end{eqnarray*}
\end{CONJECTURE}

\qq
\subsubsection{Achievable bound on performance of $\ell_1$-norm minimization} \label{subsubsec:jsm1:ell_1:achievable} \qq

\begin{figure*}[t]
\begin{center}
\epsfig{file=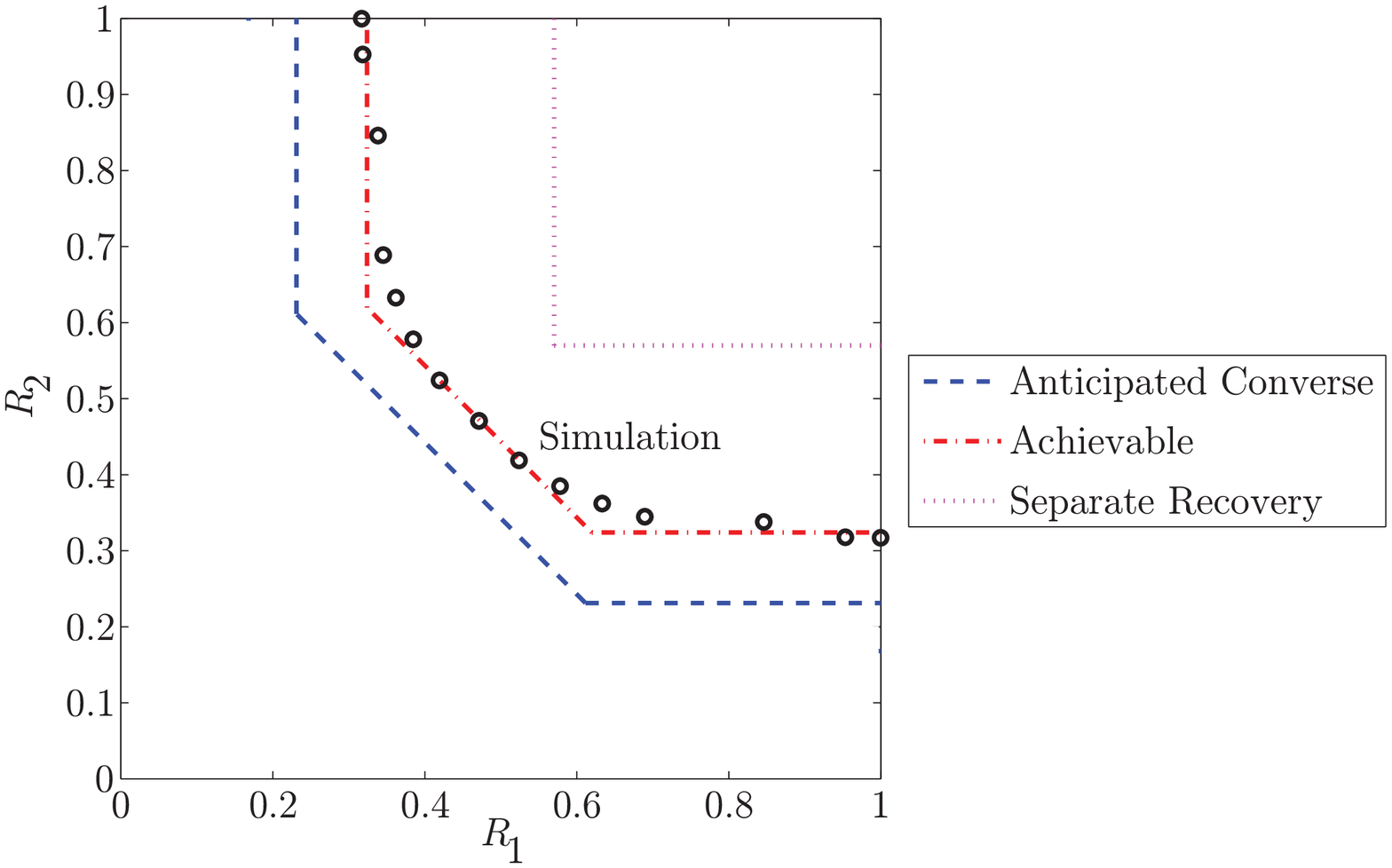,width=110mm}
\end{center}
\vspace*{-8mm}
\caption{\sl \label{fig-rateregion} Recovering a signal ensemble with
sparse common + innovations (JSM-1). We chose a
common component sparsity rate $S_C=0.2$ and innovation
sparsity rates $S_I=S_1=S_2=0.05$. Our simulation results use the
$\gamma$-weighted $\ell_1$-norm formulation (\ref{eq:gamma}) on signals
of length $\N=1000$; the measurement rate pairs that achieved perfect recovery over 100 simulations are denoted by circles.}
\end{figure*}

We have not yet characterized the performance of
$\gamma$-weighted $\ell_1$-norm formulation (\ref{eq:gamma})
analytically. Instead, Theorem~\ref{theo:achieve} below uses an alternative
$\ell_1$-norm based recovery technique. The proof describes a constructive
recovery algorithm.
We construct measurement matrices $\Phi_1$ and $\Phi_2$, each consisting of two
parts. The first parts of the matrices are identical and
recover $x_1-x_2$. The second parts of the matrices are
different and enable the recovery of
$\frac{1}{2}x_1+\frac{1}{2}x_2$. Once these two components have been
recovered, the computation of $x_1$ and $x_2$ is
straightforward. The measurement rate can be computed by considering
both identical and different parts of the measurement matrices.

\begin{THEO} \label{theo:achieve}
Let $\J=2$, $N \to \infty$ and fix the sparsity rate of the common component $S_C$
and the innovation sparsity rates $S_1=S_2=S_I$. If the measurement rates satisfy the following conditions:
\begin{subequations}
\begin{eqnarray}
\R_j &>& c'(2S_I-S_I^2),~j=1,2, \label{eqn:achieve:M1} \\
\R_1+\R_2 &>& c'(2S_I-S_I^2) +
c'(S_C+2S_I-2S_CS_I-S_I^2+S_CS_I^2), \label{eqn:achieve:sum}
\end{eqnarray}
\end{subequations}
then we can design measurement matrices
$\Phi_1$ and $\Phi_2$ with random Gaussian entries and an 
$\ell_1$-norm minimization recovery algorithm that succeeds with
probability approaching one as $N$ increases.
Furthermore, as $S_I\rightarrow{0}$ the sum measurement rate
approaches $c'(S_C)$.
\end{THEO}

The theorem is proved in Appendix~\ref{ap:achieve}.
The recovery algorithm of Theorem~\ref{theo:achieve} is based on
linear programming. It can be
extended from $\J=2$ to an arbitrary number of signals by
recovering all signal differences of the form
$x_{\j_1}-x_{\j_2}$ in the first stage of the algorithm and then
recovering $\frac{1}{\J}\sum_jx_j$ in the second stage.
In contrast, our $\gamma$-weighted $\ell_1$-norm formulation
(\ref{eq:gamma}) recovers a length-$\J\N$ signal.
Our simulation experiments (Section \ref{subsec-examplesjsm1})
indicate that the $\gamma$-weighted formulation
can recover using fewer measurements than the approach of
Theorem~\ref{theo:achieve}.

The achievable measurement rate region of Theorem~\ref{theo:achieve}
is loose with respect to the region of the anticipated converse
Conjecture~\ref{conj:ell_1_conv} (see Figure~\ref{fig-rateregion}).
We leave for future work the characterization of a tight measurement rate
region for computationally tractable (polynomial time) recovery techniques.

\qq
\subsubsection{Simulations for JSM-1} \label{subsec-examplesjsm1} \qq

We now present simulation results for several different JSM-1
settings. The $\gamma$-weighted $\ell_1$-norm formulation
(\ref{eq:gamma}) was used throughout, where the optimal choice
of $\gamma_C$, $\gamma_1$, and $\gamma_2$ depends on the relative
sparsities $\K_C$, $\K_1$, and $\K_2$. The optimal values have not been
determined analytically. Instead, we rely on a numerical optimization, which
is computationally intense. A detailed discussion of our intuition behind the
choice of $\gamma$ appears in the technical report \cite{DCSTR06}.

{\bf Recovering two signals with symmetric measurement rates:}
Our simulation setting is as follows. The signal components $z_C$,
$z_1$, and $z_2$ are assumed (without loss of generality) to be
sparse in $\Psi=I_N$ with sparsities $\K_C$, $\K_1$, and $\K_2$,
respectively. We assign random Gaussian values to the nonzero
coefficients. We restrict our attention to the symmetric setting in
which $\K_1=\K_2$ and $\M_1 = \M_2$, and consider signals of length
$\N=50$ where $\K_C+\K_1+\K_2=15$.

\begin{figure*}[t]
\begin{center}
\begin{tabular}{ccc}
\epsfig{file=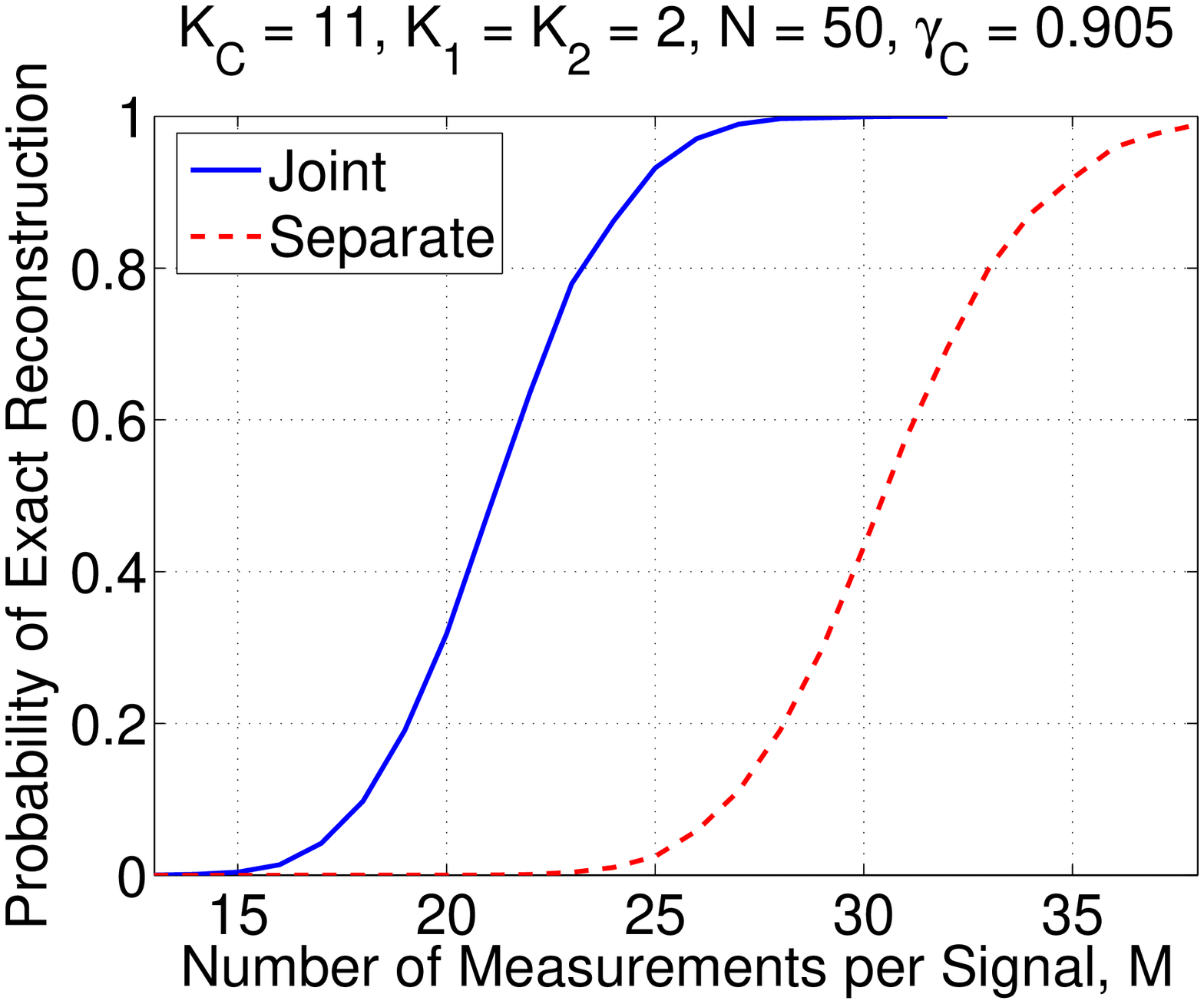,height=55mm} &
\epsfig{file=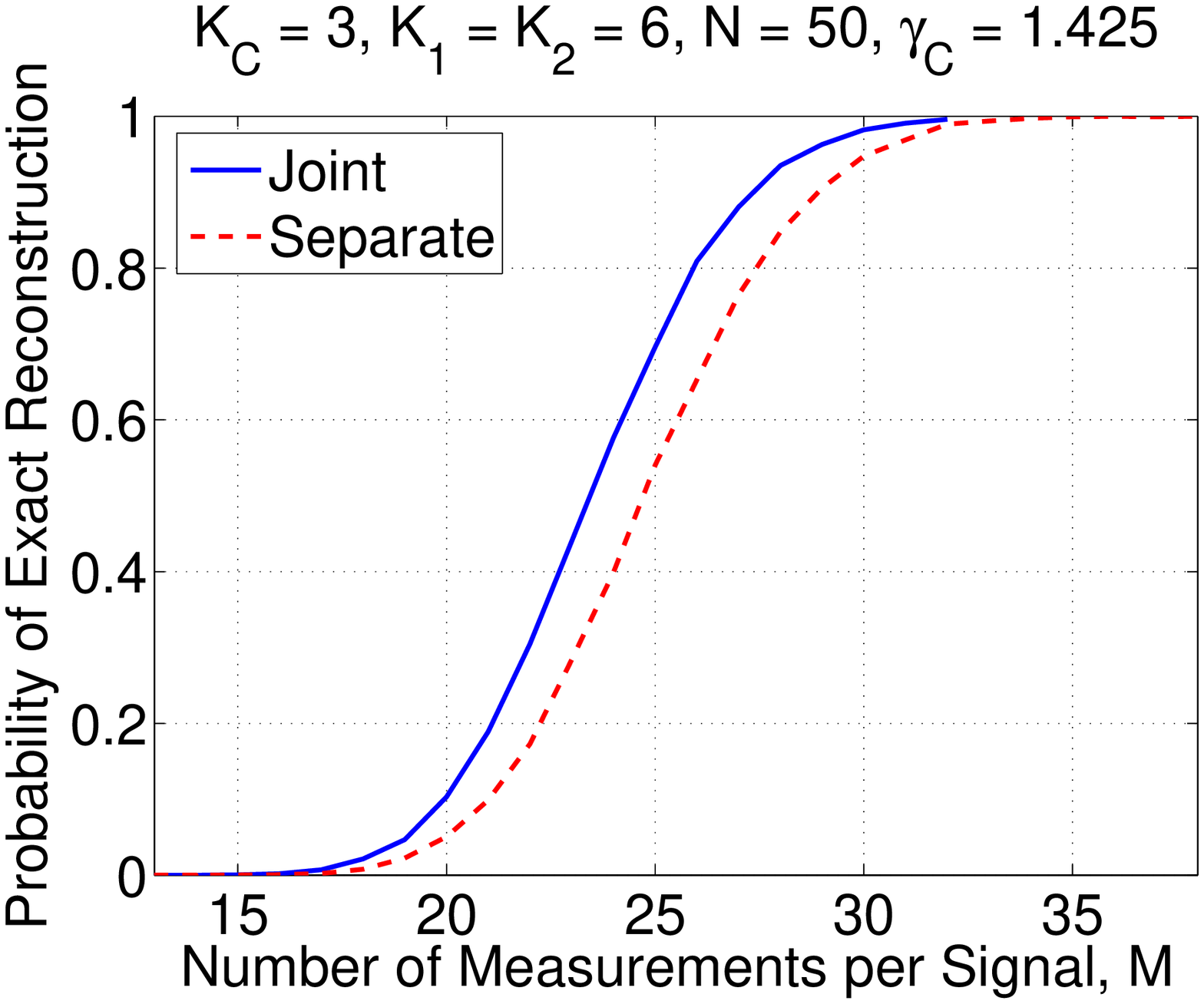,height=55mm}
\end{tabular}
\end{center}
\vspace*{-6mm}
\caption{\sl \label{fig-symmetric} Comparison of joint decoding and
separate decoding for JSM-1. The advantage of joint over separate
decoding depends on the common component sparsity. }
\end{figure*}

In our joint decoding simulations, we consider values of $\M_1$ and
$\M_2$ in the range between 10 and 40. We find the optimal
$\gamma_C$ in the $\gamma$-weighted $\ell_1$-norm formulation
(\ref{eq:gamma}) using a line search optimization, where simulation
indicates the ``goodness" of specific $\gamma_C$ values in terms of
the likelihood of recovery. 
With the optimal $\gamma_C$, for
each set of values we run several thousand trials to determine the
empirical probability of success in decoding $z_1$ and $z_2$. The
results of the simulation are summarized in
Figure~\ref{fig-symmetric}. The savings in
the number of measurements $\M$ can be substantial,
especially when the common component $\K_C$ is large
(Figure~\ref{fig-symmetric}). For $\K_C=11$, $\K_1=\K_2=2$, $\M$ is
reduced  by approximately $30\%$. For smaller $\K_C$, joint decoding
barely outperforms separate decoding, since most of the measurements
are expended on innovation components. Additional results appear in
\cite{DCSTR06}.

{\bf Recovering two signals with asymmetric measurement rates:}
In Figure~\ref{fig-rateregion}, we compare separate CS
recovery with the anticipated converse bound of Conjecture~\ref{conj:ell_1_conv},
the achievable bound of Theorem~\ref{theo:achieve}, and numerical results.

We use $\J=2$ signals and choose a common component sparsity rate
$S_C=0.2$ and innovation sparsity rates $S_I=S_1=S_2=0.05$. We consider 
several different asymmetric measurement rates. In each such setting, we
constrain $\M_2$ to have the form $\M_2=\alpha \M_1$ for some
$\alpha$, with $\N=1000$. The
results plotted indicate the smallest pairs $(\M_1,\M_2)$ for which
we always succeeded recovering the signal over $100$ simulation
runs. In some areas of the measurement
rate region our $\gamma$-weighted $\ell_1$-norm formulation (\ref{eq:gamma})
requires fewer measurements than the achievable approach of
Theorem~\ref{theo:achieve}.

{\bf Recovering multiple signals with symmetric measurement
rates:} The $\gamma$-weighted $\ell_1$-norm recovery technique of this
section is especially promising when $\J>2$ sensors are used.
These savings may be valuable in applications such as sensor networks, where
data may contain strong spatial (inter-source) correlations.

\begin{figure*}[t]
\begin{center}
\epsfig{file=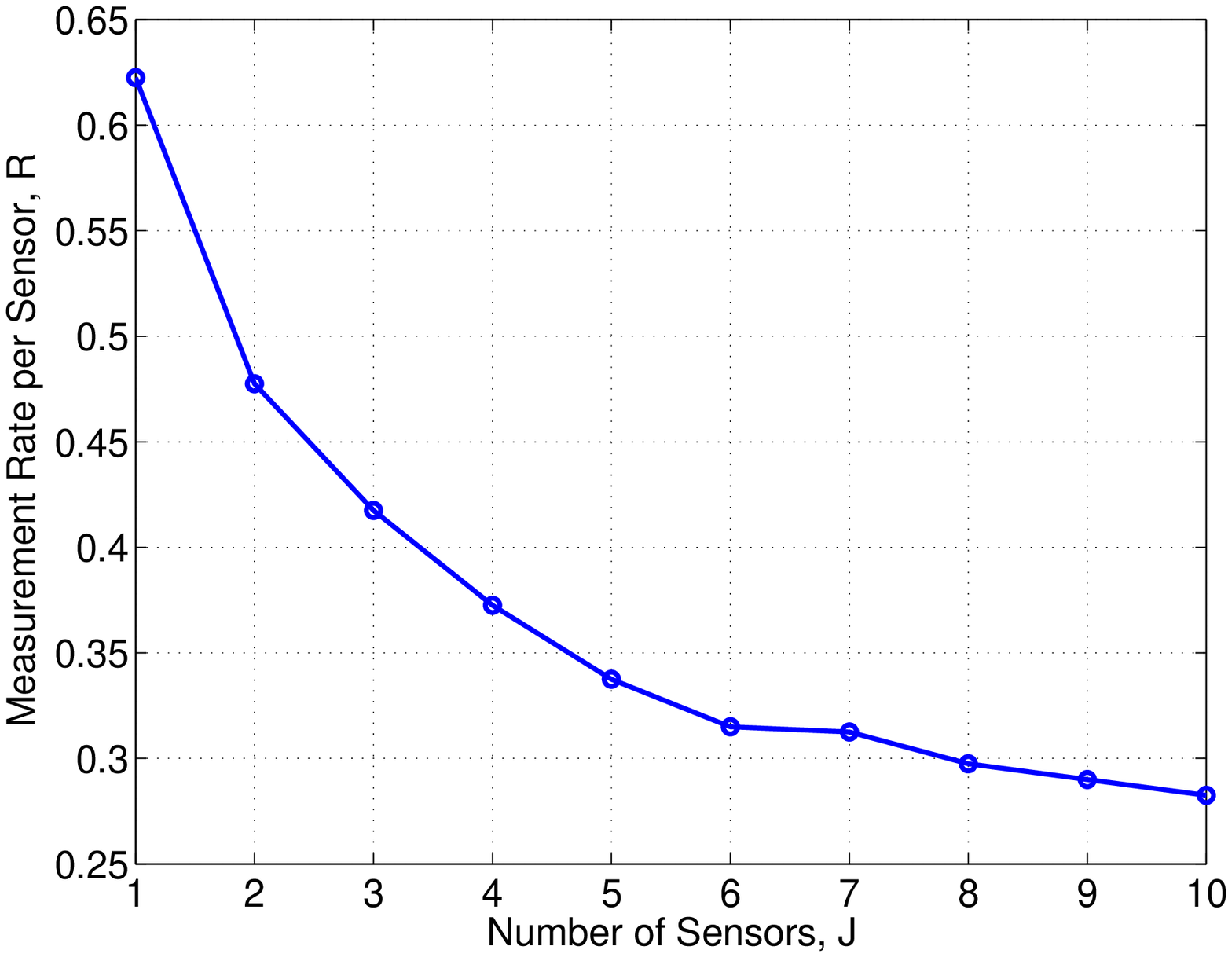,height=60mm}
\end{center}
\vspace*{-8mm}
\caption{\sl \label{fig:multisensor} Multi-sensor measurement
results for JSM-1. We choose a common component sparsity rate $S_C=0.2$,
innovation sparsity rates $S_I=0.05$, and signals of length
$\N=500$; our results demonstrate a reduction in the measurement rate per 
sensor as the number of sensors $J$ increases. }
\end{figure*}

We use $\J\in\{1, 2,\ldots,10\}$ signals and choose the same sparsity
rates $S_C=0.2$ and $S_I=0.05$ as the asymmetric rate simulations;
here we use symmetric measurement rates and let $\N=500$.
The results of Figure~\ref{fig:multisensor} describe the smallest
symmetric measurement rates for which we always succeeded
recovering the signal over $100$ simulation runs. As
$\J$ increases, lower measurement rates can be used; the results compare
favorably with the lower bound from Conjecture~\ref{conj:ell_1_conv}, which
gives $R_j \approx 0.232$ as $J \to \infty$.

%% file: jsm1proofs.tex
\qq
\section{Proof of Lemma~\ref{lem:con}}
\label{ap:lem_con} \qq

{\bf Necessary conditions on innovation components:}\ We begin by
proving that in order to recover $z_C$,
$z_1$, and $z_2$ via the $\gamma$-weighted
$\ell_1$-norm formulation it is necessary that $z_1$ can be
recovered via single-signal $\ell_1$-norm minimization using
$\Phi_1$ and measurements $\bar{y_1} = \Phi_1z_1$.

Consider the single-signal $\ell_1$-norm minimization problem
\begin{equation*}
\bar{z_1} = \arg\min \|z_1\|_1 ~~~\mbox{s.t. }
\bar{y_1}=\Phi_1z_1.
\end{equation*}
Suppose that this $\ell_1$-norm minimization for $z_1$
fails; that is, there exists $\bar{z_1}\neq z_1$
such that $\bar{y_1}=\Phi_1 \bar{z_1}$ and
$\|\bar{z_1}\|_1\leq \|z_1\|_1$. Therefore,
substituting $\bar{z_1}$ instead of $z_1$ in the
$\gamma$-weighted $\ell_1$-norm formulation (\ref{eq:gamma}) provides an
alternate explanation for the measurements with a smaller or equal
modified $\ell_1$-norm penalty. Consequently, recovery of
$z_1$ using (\ref{eq:gamma}) will fail and we will
recover $x_1$ incorrectly. We conclude that the single-signal
$\ell_1$-norm minimization of $z_1$ using $\Phi_1$ is
necessary for successful recovery using the $\gamma$-weighted
$\ell_1$-norm formulation. A similar condition for $\ell_1$-norm minimization of $z_2$ using $\Phi_2$ and measurements
$\Phi_2z_2$ can be proved in an analogous manner.

{\bf Necessary condition on common component:}\ We now prove that in
order to recover $z_C$, $z_1$, and
$z_2$ via the $\gamma$-weighted $\ell_1$-norm formulation it
is necessary that $z_C$ can be recovered via
single-signal $\ell_1$-norm minimization using the joint matrix
$[\Phi_1\trans~~\Phi_2\trans]\trans$ and measurements
$[\Phi_1\trans~~\Phi_2\trans]\trans z_C$.

The proof is very similar to the previous proof for the innovation
component $z_1$. Consider the single-signal $\ell_1$-norm
minimization
\begin{equation*}
\bar{z}_C = \arg\min \|z_C\|_1 ~~~\mbox{s.t. }
\bar{y}_C=[\Phi_1\trans~~\Phi_2\trans]\trans z_C.
\end{equation*}
Suppose that this $\ell_1$-norm minimization for $z_C$
fails; that is, there exists $\bar{z}_C\neq z_C$
such that
$\bar{y}_C=[\Phi_1\trans~~\Phi_2\trans]\trans\bar{z}_C$
and $\|\bar{z}_C\|_1\leq \|z_C\|_1$. Therefore,
substituting $\bar{z}_C$ instead of $z_C$ in the
$\gamma$-weighted $\ell_1$-norm formulation (\ref{eq:gamma}) provides an
alternate explanation for the measurements with a smaller modified
$\ell_1$-norm penalty. Consequently, the recovery of
$z_C$ using the $\gamma$-weighted $\ell_1$-norm formulation
(\ref{eq:gamma}) will fail, and thus we will recover $x_1$ and
$x_2$ incorrectly. We conclude that the single-signal $\ell_1$-norm minimization of $z_C$ using $[\Phi_1\trans~~\Phi_2\trans]\trans$ is necessary for successful
recovery using the $\gamma$-weighted $\ell_1$-norm formulation.
\qed

\qq
\section{Proof of Theorem~\ref{theo:achieve}}
\label{ap:achieve} \qq

We construct measurement matrices $\Phi_1$ and $\Phi_2$ that consist
of two sets of rows. The first set of rows is identical in both and
recovers the signal {\em difference} $x_1-x_2$. The second set
is different and recovers the signal {\em average}
$\frac{1}{2}x_1+\frac{1}{2}x_2$. Let the submatrix formed by the
identical rows for the signal difference be $\Phi_D$, and let the
submatrices formed by unique rows for the signal average be
$\Phi_{A,1}$ and $\Phi_{A,2}$. Thus the measurement
matrices $\Phi_1$ and $\Phi_2$ are of the following form:
\begin{equation*}
\Phi_1 = \left[
\begin{array}{ccc}
\Phi_D \\
\Phi_{A,1} \end{array} \right] \mbox{    and     } \Phi_2 = \left[
\begin{array}{ccc}
\Phi_D \\
\Phi_{A,2} \end{array} \right].
\end{equation*}
The submatrices $\Phi_D$, $\Phi_{A,1}$, and $\Phi_{A,2}$ contain
i.i.d.\ Gaussian entries. Once the difference $x_1-x_2$ and average
$\frac{1}{2}x_1+\frac{1}{2}x_2$ have been recovered using the
above technique, the computation of $x_1$ and $x_2$ is
straightforward. The measurement rate can be computed by considering
both parts of the measurement matrices.

{\bf Recovery of signal difference:}\ The submatrix $\Phi_D$
is used to recover the signal difference. By subtracting the
product of $\Phi_D$ with the signals $x_1$ and $x_2$, we have
\begin{equation*}
\Phi_D x_1 - \Phi_D x_2 = \Phi_D (x_1-x_2).
\end{equation*}
In the original representation we have $x_1-x_2=z_1-z_2$ with
sparsity rate $2S_I$. But $z_1(n)-z_2(n)$ is nonzero only if
$z_1(n)$ is nonzero or $z_2(n)$ is nonzero. Therefore, the sparsity
rate of $x_1-x_2$ is equal to the sum of the individual sparsities
reduced by the sparsity rate of the overlap, and so we have
$S(X_1-X_2)=2S_I-(S_I)^2$. Therefore, any measurement rate greater
than $c'(2S_I-(S_I)^2)$ for each $\Phi_D$ permits recovery of
the length $\N$ signal $x_1-x_2$. (As always, the probability of
correct recovery approaches one as $\N$ increases.)

{\bf Recovery of average:}\ Once $x_1-x_2$ has been
recovered, we have
\begin{equation*}
x_1-\frac{1}{2}(x_1-x_2) = \frac{1}{2} x_1+\frac{1}{2} x_2 =
x_2+\frac{1}{2}(x_1-x_2).
\end{equation*}
At this stage, we know $x_1-x_2$, $\Phi_D x_1$, $\Phi_D x_2$,
$\Phi_{A,1} x_1$, and $\Phi_{A,2} x_2$. We have
\begin{eqnarray*}
\Phi_D x_1 - \frac{1}{2} \Phi_D (x_1-x_2) = \Phi_D
\left(\frac{1}{2}x_1+\frac{1}{2}x_2 \right), \\
\Phi_{A,1} x_1 - \frac{1}{2} \Phi_{A,1} (x_1-x_2) = \Phi_{A,1}
\left(\frac{1}{2}x_1+\frac{1}{2}x_2 \right), \\
\Phi_{A,2} x_2 + \frac{1}{2} \Phi_{A,2} (x_1-x_2) = \Phi_{A,2}
\left(\frac{1}{2}x_1+\frac{1}{2}x_2 \right),
\end{eqnarray*}
where $\Phi_D (x_1-x_2)$, $\Phi_{A,1} (x_1-x_2)$, and $\Phi_{A,2}
(x_1-x_2)$ are easily computable because $(x_1-x_2)$ has been
recovered. The signal $\frac{1}{2}x_1+\frac{1}{2}x_2$ is of
length $\N$;  its sparsity rate is equal to the sum
of the individual sparsities $S_C+2S_I$ reduced by the sparsity rate
of the overlaps, and so we have
$S(\frac{1}{2}X_1+\frac{1}{2}X_2)=S_C+2S_I-2S_CS_I-(S_I)^2+S_C(S_I)^2$.
Therefore, any measurement rate greater than
$c'(S_C+2S_I-2S_CS_I-(S_I)^2+S_C(S_I)^2)$ aggregated over the
matrices $\Phi_D$, $\Phi_{A,1}$, and $\Phi_{A,2}$ enables
recovery of $\frac{1}{2}x_1+\frac{1}{2}x_2$.

{\bf Computation of measurement rate:}\ By considering the
requirements on $\Phi_D$, the individual measurement rates $\R_1$
and $\R_2$ must satisfy (\ref{eqn:achieve:M1}). Combining the measurement
rates required for $\Phi_{A,1}$ and $\Phi_{A,2}$, the sum
measurement rate satisfies (\ref{eqn:achieve:sum}). We complete the
proof by noting that $c'(\cdot)$ is continuous and that
$\lim_{S\rightarrow{0}}c'(S)=0$. Thus, as $S_I$ goes to zero, the limit of the sum
measurement rate is $c'(S)$. \qed